\documentclass[aps,prl,nobalancelastpage,twocolumn,reprint]{revtex4-2}
\usepackage[utf8]{inputenc}
\usepackage[T1]{fontenc}
\usepackage[caption=false]{subfig}
\usepackage{times}
\usepackage{amsmath}
\usepackage{amsfonts}
\usepackage{amssymb}
\usepackage{graphicx}
\usepackage{listings}
\usepackage{hyperref}
\usepackage{bm}
\usepackage{braket}
\usepackage{import}
\usepackage{siunitx}
\usepackage{pdfpages}
\usepackage{tikz}
\usepackage[normalem]{ulem}

\makeatletter
\AtBeginDocument{\let\LS@rot\@undefined}
\makeatother

\begin{document}

\begin{abstract}
   Active solids such as cell collectives, colloidal clusters, and active metamaterials exhibit diverse collective phenomena, ranging from rigid body motion to shape-changing mechanisms. The nonlinear dynamics of such active materials remains however poorly understood
   when they host zero-energy deformation modes and when noise is present. Here, we show that stress propagation in a model of active solids induces 
   the spontaneous actuation of multiple soft floppy modes, even without exciting vibrational modes.
  By introducing an adiabatic approximation, we map the dynamics onto an effective Landau free energy, predicting mode selection and the onset of collective dynamics.
   These results open new ways to study and design living and robotic materials with multiple modes of locomotion and shape-change.
\end{abstract}

\title{Active Solids Model: Rigid Body Motion and Shape-changing Mechanisms}
\author{Claudio Hern\'andez-L\'opez$^{1,4}$ }
\author{Paul Baconnier$^2$ }
\author{Corentin Coulais$^3$}
\author{Olivier Dauchot$^2$ }
\author{Gustavo Düring$^4$}
\affiliation {$^1$ Laboratoire de Physique de l’\'Ecole Normale Sup\'erieure,  UMR CNRS 8023, Université PSL, Sorbonne Université, 75005 Paris, France}
\affiliation{$^2$ Gulliver UMR CNRS 7083, ESPCI Paris, Universit\'e PSL, Paris, France}
\affiliation{$^3$Institute of Physics, Universiteit van Amsterdam, 1098 XH Amsterdam, The Netherlands}
\affiliation{$^4$Instituto de Física, Pontificia Universidad Católica de Chile, 8331150 Santiago, Chile}

\maketitle
Polar active matter is composed of self-driven units that convert energy into directed motion or forces. Aligning interactions among the active units lead to large scale collective motion in various forms, from polar flocks of birds~\cite{Cavagna-2010,bialek_statistical_2012}, motile colloids~\cite{bricard_emergence_2013}, vibrated disks~\cite{weber_long-range_2013}, interacting robots~\cite{ferrante_elasticity-based_2013, zheng_experimental_2020,D1SM00080B}, and vortex flows of fish~\cite{flierl_individuals_1999}, bacteria~\cite{Liu-2021}, or colloids~\cite{Bricard-2015,Chardac-2021}, to only quote a few.
The large scale physics of these flows has been the topic of intensive research and is well described by the so-called Toner-Tu equations~\cite{Toner-2005, Toner-2012}. 
When the density of active units is large because of confinement~\cite{henkes_active_2011} or cohesion~\cite{shimoyama_collective_1996,Gregoire-2004}, the structure of the assembly may remain frozen on long time scales, and the system exhibits elastic rather than viscous properties. When cohesive interactions are large enough, as is the case for dense biofilms~\cite{Xu-2023}, keratocyte swarms~\cite{szabo_phase_2006} or Epithelial monolayers \cite{peyret_sustained_2019}, active units can be considered embedded in an elastic network, in a way similar to artificially designed active elastic metamaterials~\cite{Ferrante-2013a,Woodhouse-2018,baconnier_selective_2022,Zheng-2023}.   

A natural starting point is then to analyze the dynamics in terms of the vibrational modes of the elastic medium or structure, with the zero modes corresponding to node displacements that do not change any bond length \cite{Xiaoming2018}. It was shown that correlated noise generated by an active matter bath can actuate a non-trivial zero mode while suppressing harmonic modes to a degree dependent on temporal correlations~\cite{Woodhouse-2018}.
Self-propulsion is further able to mobilize solid body motion~\cite{Ferrante-2013a} or a free-moving mechanism even in a topologically complex case~\cite{Woodhouse-2018}. 
Notably, observations on \textit{Placozoa phylum}~\cite{Couzin2023}, a living active solid, have revealed global rotation and translations under various conditions. Finally, in the presence of a non-linear feedback of the elastic stress on the orientation of the active forces, self-propulsion can also actuate a few selected harmonic modes~\cite{baconnier_selective_2022}. %
Yet the selection mechanism remains unclear. More generally, in the presence of several actuatable modes, whether trivially associated with solid body motion or more complex mechanisms, several dynamics coexist in phase space and there is to date no general principle to characterize their metastability.

\begin{figure}[t]
\centering
{\includegraphics[width=.75\linewidth]{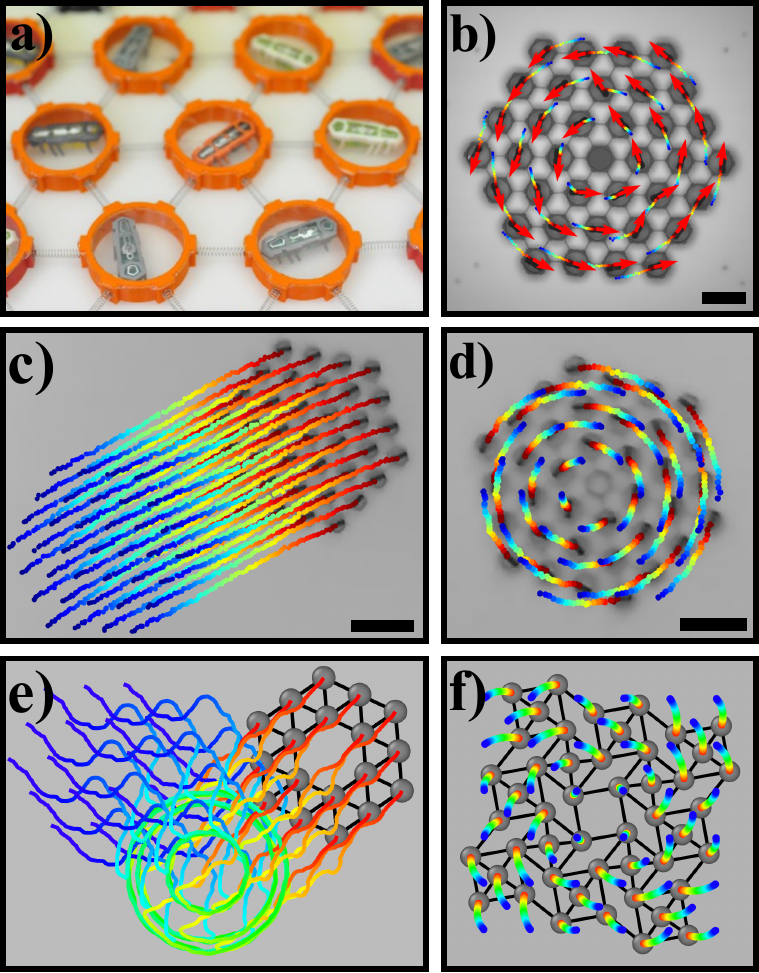}}
\caption{\textbf{Active rigid body motion and active mechanisms} 
(a) Zoom on the experimental active elastic lattice introduced in~\cite{baconnier_selective_2022}, with self-propelling units - Hexbugs - trapped in 3d-printed annuli, connected by springs in a triangular lattice.
(b) Experimental rotational dynamics observed under central pinning.  Scale bar: 10 cm.
(c,d) Experimental translational and rotational dynamics observed for a free structure. Scale bar: 20 cm.
(e) Alternating translational and rotation dynamics obtained numerically for the same free structure.
(f) A rotational-auxetic regime observed numerically for a non-pinned auxetic square system (See text, Fig.~\ref{fig:fig3}, and Sup. Movie 1 \cite{SM}). Trajectories color-coded from blue to red by increasing time.
}
\label{fig:principle}
\vspace{-5mm}
\end{figure}

\begin{figure*}
\centering
\includegraphics[width=0.975\linewidth]{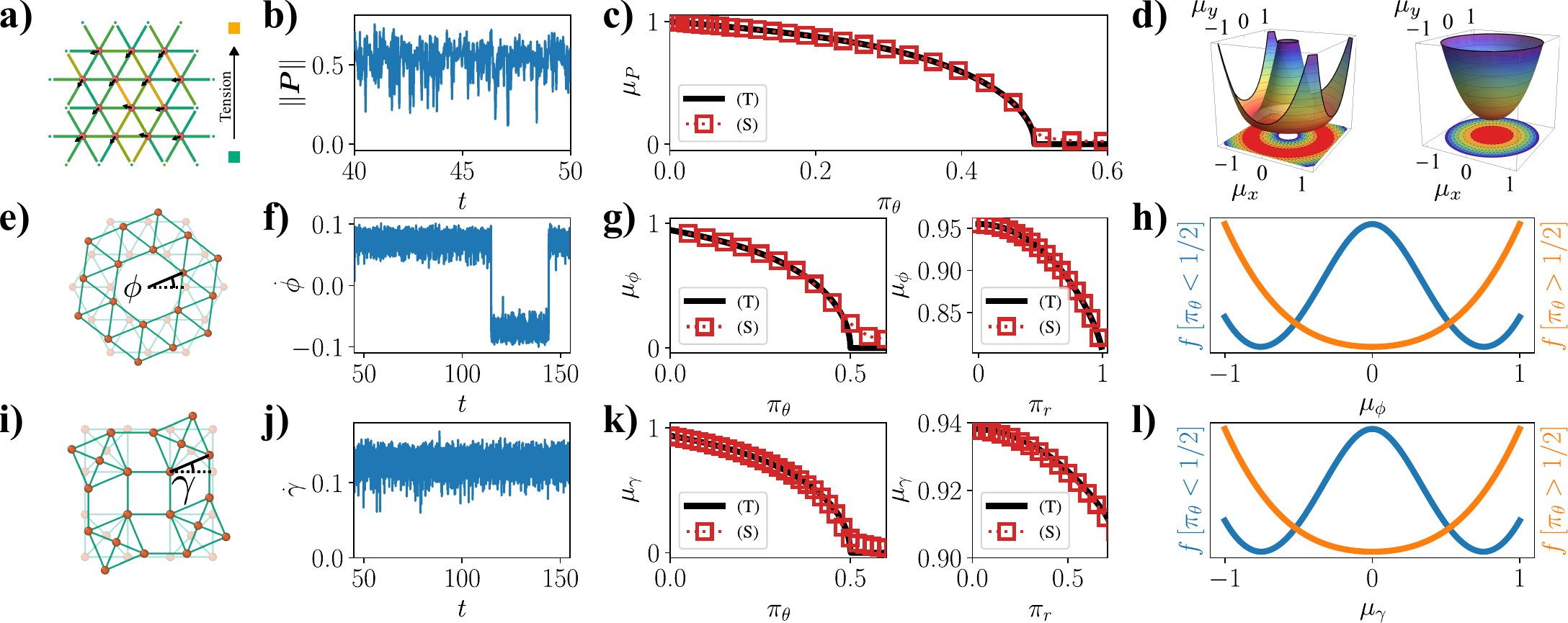}
\caption{
\small{\textbf{Second order transitions to solid body motion and mechanisms for 1-mode systems}}
Simulations consider $\pi_r = 0.001$ unless specified.
(a) Periodic boundary condition (translation only) network. (b-d) consider a 100x100 translating system.
(b) Global polarization magnitude $\lVert \boldsymbol{P} \rVert$ time series; $\pi_{\theta} = 0.4$.
(c) Phase diagram $\mu_P$ vs. $\pi_\theta$ as a function of noise. (S) for simulations, (T) for theoretical predictions.
(d) Landau free energy as a function of 
$\mu_{x}$ and $\mu_{y}$;
left: $\pi_{\theta}=0.1$; right: $\pi_{\theta}=0.6$.
(e) A 2-layer hexagonal ring with the definition of the rotational angle $\phi$. (f-h) consider a 9-layer hexagonal ring system.
(f) Angular velocity $\dot{\phi}$ time series; $ \pi_{\theta} = 0.4$.
(g) Phase diagram $\mu_\phi$ vs. $\pi_\theta$ (left) and $\mu_\phi$ vs. $\pi_r$, when $\pi_\theta=0$ (right).
(h) Landau free energy as a function of $\mu_\phi$; blue: $\pi_{\theta}=0.25$; orange: $\pi_{\theta}=0.6$.
(i) A 1-layer auxetic system with the definition of the auxetic angle $\gamma$. (j-l) consider an 8-layer auxetic system.
(j) Auxetic angular velocity $\dot{\gamma}$ time series; $\pi_{\theta} = 0.4$.
(k) Phase diagram $\mu_\gamma$ vs. $\pi_\theta$ (left) and $\mu_\gamma$ vs. $\pi_r$, when $\pi_\theta=0$ (right).
(l) Landau free energy as a function of $\mu_\gamma$.
blue: $\pi_{\theta}=0.25$; orange: $\pi_{\theta}=0.6$.}
\label{fig:fig2}
\end{figure*}

In this Letter, we provide a general formalism to describe the statistical evolution of collective motion, in the case where several zero modes are present, as illustrated in Fig.~\ref{fig:principle}, using the hexbug elastic network (Fig.~\ref{fig:principle}-a), introduced in~\cite{baconnier_selective_2022}.  
When the network is pinned in the center and the translational solid body motion is forbidden, the only remaining zero mode is rotation (Fig.~\ref{fig:principle}-b). The dynamics breaks chiral symmetry by spontaneously selecting one direction of rotation, which eventually reverses in the presence of noise. When the network is not pinned, there are two translational and one rotational zero modes, which also spontaneously break the continuous rotational, XY model-like, and chiral, Ising-like, symmetries. Both translational (Fig.~\ref{fig:principle}-c) and rotational motion (Fig.~\ref{fig:principle}-d) are observed depending on the initial conditions. Transitions between the two types of motion are observed numerically (Fig.~\ref{fig:principle}-e). Finally, Fig.~\ref{fig:principle}-f, shows the dynamics of a non-trivial active mechanism with two zero modes: an ideal auxetic network~\cite{resch1965geometrical,grima2000auxetic,Acuna2022,Czajkowski-2022} 
pinned at the center that can freely rotate and contract. 

When the timescales of the dynamics are much longer than the elastic relaxation time, as is the case here, the harmonic modes of the solid are barely excited and the network can be considered as rigid. 
We show first that in this limit stress propagation is enough to induce collective motion: at small enough noise, the symmetric phase is spontaneously broken and 
the evolution of the system
follows a specific path along the zero mode space. This introduces a new timescale, which competes with the timescale of reorientation of the active particles. We prove that within an adiabatic approximation, the dynamics is governed by an effective Landau free energy, from which the mode selection and the metastability 
can be easily understood. Our results pave the way towards active metamaterials with multiple modes of actuation and locomotion \cite{brandenbourger_limit_2021}.

We consider active systems described by the overdamped dynamics of $N$ self-aligning units, which were introduced independently in several contexts ~\cite{shimoyama_collective_1996,szabo_phase_2006,henkes_active_2011,weber_long-range_2013,dauchot_dynamics_2019, baconnier_selective_2022}.
Written in non-dimensional units (See Sup. Mat. \cite{SM}, section I for details), the equations read:
\begin{align}
	\dot{\boldsymbol{x}}_i &=  \boldsymbol{\hat{n}}_i + \boldsymbol{F}_i, \label{eq:dimensionless_x} \\
	 \dot{\theta}_i &= \frac{1}{\pi_r} (\boldsymbol{\hat{n}^\perp}_i \cdot \boldsymbol{F}_i)
  + \sqrt{2 \pi_{\theta} / \pi_r} \xi_i, \label{eq:dimensionless_n}
\end{align}
with $\boldsymbol{x}_i$, respectively $\boldsymbol{\hat{n}}_i = (\cos\theta_i, \sin\theta_i)$, the position and the polarization unit vector of active unit $i$. 
$\boldsymbol{F}_i$ is the net external force on unit $i$.
The dimensionless self-alignment length $\pi_r = \ell_a/\ell_0$ is the ratio between the self-alignment length $\ell_a$ and the characteristic agent-agent distance $\ell_0$. The dimensionless noise coefficient corresponds to $\pi_{\theta} = D_{\theta} \ell_a/v_0$, with $v_0$ the speed of a free agent, and $D_{\theta}$ the angular diffusion coefficient. We define $\xi_i$ as a delta-correlated Gaussian white noise process.
The self-aligning torque, on the right-hand side of Eq.~\ref{eq:dimensionless_n}, emerges from non-symmetric dissipative forces with respect to $\boldsymbol{\hat{n}}_i$, when it is misaligned with $\dot{\boldsymbol{x}} _i$~\cite{dauchot_dynamics_2019}. It was shown to be the key ingredient for the onset of collective motion in active disks \cite{weber_long-range_2013}, and collective actuation in active elastic networks~\cite{baconnier_selective_2022}. 

One can show (See Sup. Mat. \cite{SM}, section I) that the network can be safely considered rigid provided
\begin{align}
\frac{\ell_e}{\ell_a} \ll \frac{1}{k} \omega^2_q, \quad \forall q \in \{1..,2N\}
\label{rigidcond}
\end{align}
where $\ell_e$ is the amplitude of the typical displacement of the nodes resulting from the active forces, 
and  $\omega_q^2$ are the non-zero eigenvalues of the dynamical matrix \cite{Xiaoming2018}. As such, the bound for activation of the elastic modes only depends on the geometry of the structure. In the following, we shall perform all simulations with parameters such that the rigid approximation holds. 

We start by simulating three systems with a single zero mode (Fig.~\ref{fig:fig2}):  (i) a crystalline triangular lattice with periodic boundary conditions (PBC), (ii) a triangular lattice pinned at its center, and (iii) an ideal auxetic network pinned at its center, illustrated in Fig.~\ref{fig:fig2}-a,e,i respectively. All of them exhibit collective motion along their single zero mode at small noise. The triangular lattice with PBC translates uniformly, with a non-zero magnitude of the global polarization $\boldsymbol{P} = (1/N) \sum_i \boldsymbol{\hat{n}}_i$ (Fig.~\ref{fig:fig2}-b). The network pinned at the center freely rotates, with an angular speed $\dot\phi$, that randomly switches from counterclockwise to clockwise rotation (Fig.~\ref{fig:fig2}-f). The auxetic network freely compresses with a finite auxetic angular speed $\dot\gamma$ (Fig.~\ref{fig:fig2}-j). In all cases, collective motion emerges from a spontaneous symmetry breaking of the disordered phase, when $\pi_\theta < 1/2$ (Fig.~\ref{fig:fig2}-c,g,k). 

\begin{figure*}
\centering
\includegraphics[width=0.975\linewidth]{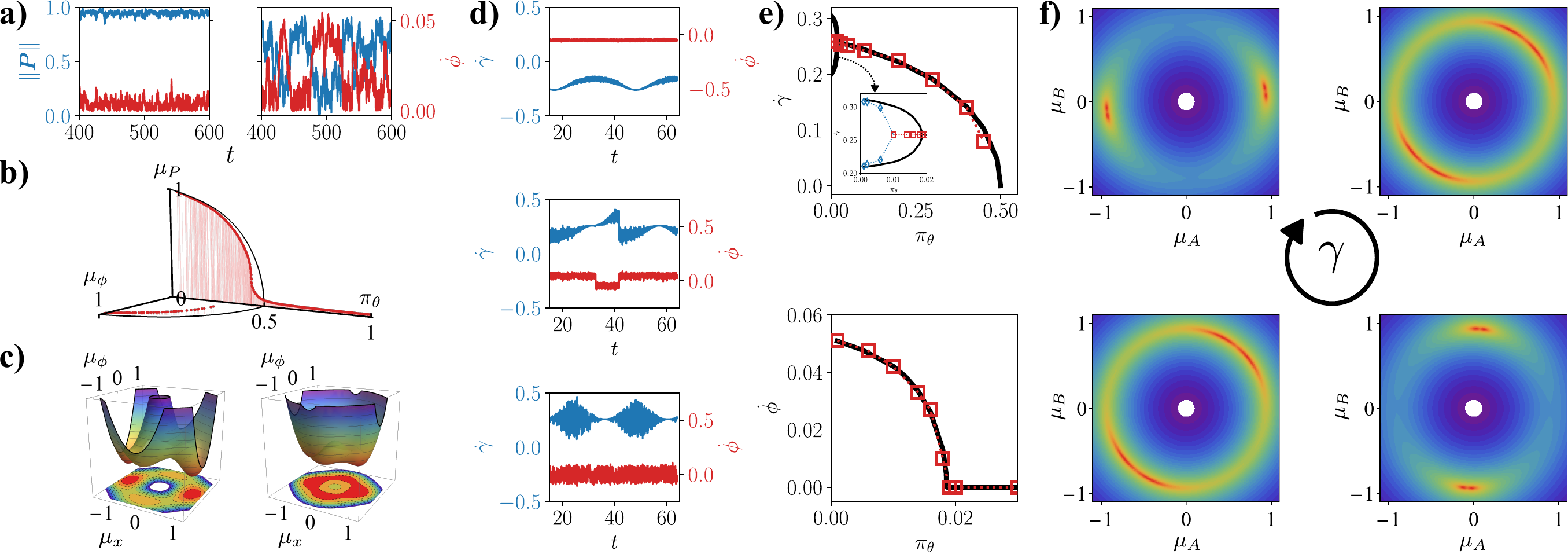}
\caption{ \small{\textbf{2-mode systems phenomenology: switching between modes of actuation.}
(a) Time series of the magnitude of the global polarization $\boldsymbol{P}$ (blue) and of the angular velocity $\dot{\phi}$ (red) for a 2-ring non pinned triangular lattice; left to right: $\pi_{\theta} = 0.10, 0.35$; $\pi_{r} = 0.1$.
(b) Phase diagram $\mu_{P}$ and $\mu_{\phi}$ vs. $\pi_\theta$ of a 30-ring non pinned triangular lattice.
(c) Landau free energy for a non-pinned 2-ring triangular lattice as a function of $\mu_\phi$ and $\mu_x$; from left to right: $\pi_{\theta} = 0.10, 0.35$. (d-f) consider an 8-layer rotational-auxetic network.
(d) Time series of the auxetic angular velocity $\dot{\gamma}$ (blue) and rotational angular velocity $\dot{\phi}$ (red);  from top to bottom: $\pi_{\theta} = 0.001, 0.006, 0.014$.
(e) Phase diagram $\mu_\gamma$ (top) and $\mu_\phi$ (bottom) vs. $\pi_\theta$. The theoretical values are the time average of each solution obtained from following the free energy minima as $\gamma$ varies. Inset: zoom-in to the $\dot{\phi} \neq 0$ region.
(f) Landau free energy as it varies with $\gamma$ ($\gamma = 0.2 + n \pi/2$, with $n=0,1,2,3$); $\pi_{\theta}=0.001$, and the evolution time step $\Delta t = 0.01$.}}
\label{fig:fig3}
\end{figure*}

The situation becomes more interesting when the network of interest has more than one zero mode, see Fig. \ref{fig:fig3}. An active network free of PBC has both translational and rotational zero modes. 
The simulations reveal that, at low $\pi_\theta$, the collective dynamics switches between pure translations and pure rotation 
(Fig.~\ref{fig:fig3}-a). The case of an auxetic network that is also free to rotate is even more complex. As we shall see this is because the rotational mode depends on the distance of the particles to the center, 
which changes while the system evolves along the auxetic mode. 
The visual inspection of the auxetic and rotation rates as a function of time for different values of $\pi_\theta$ indicates that in the limit of vanishing noise, $\dot{\phi}$ remains 
constant, and $\dot{\gamma}$ fluctuates periodically. Increasing the noise, transitions between two states with different
$\dot{\phi}$ signs can be achieved, and even larger noise values lead to a state where such transitions occur constantly (Fig.~\ref{fig:fig3}-d). 

We now come to the theoretical analysis of the above observations. 
For a rigid network, the bond elongations are null.  Imposing such distance-preserving condition to Eq.~\ref{eq:dimensionless_x} one finds after some algebra (See Sup. Mat. \cite{SM}, section II) :
\begin{equation}
\label{ForceR}
	\boldsymbol{F}_i = - \boldsymbol{\hat{n}}_i + \sum_{q \in \mathfrak{F}} 
	\langle \boldsymbol{\varphi}^q|  \boldsymbol{\hat{n}}\rangle \boldsymbol{\varphi}_i^q,
\end{equation}
where $\boldsymbol{\varphi}_i^q$ is the vector associated with particle $i$ in the $q$-th eigenmode of the dynamical matrix of the elastic network, and $\mathfrak{F}$ is the set of zero modes. Note that self-stress states, i.e stress configurations in the null space of the equilibrium matrix \cite{Xiaoming2018} do not contribute to this net force, and thus do not influence the behavior of the system.
Then, replacing the force in Equations (\ref{eq:dimensionless_x}) and (\ref{eq:dimensionless_n}): 
\begin{align}
	\dot{\boldsymbol{x}}_i =   \sum_{q \in \mathfrak{F}} 
	\langle \boldsymbol{\varphi}^q|  \boldsymbol{\hat{n}}\rangle \boldsymbol{\varphi}_i^q
\quad\text{and} \quad\dot{\theta}_i = -\frac{1}{\pi_r} \frac{\partial V}{\partial \theta_i}  + \sqrt{\frac{2 \pi_{\theta}}{\pi_r}}\xi_i,
	     \label{eq:modes_n}
\end{align}
with $V= -\frac{1}{2} \sum_{q \in \mathfrak{F}} \langle \boldsymbol{\varphi}^q | \boldsymbol{\hat{n}}\rangle^2.$ 
The stochastic dynamics is then described by the time-dependent probability density $Q(\boldsymbol{x}_1,\dots,\boldsymbol{x}_N; \theta_1, \dots, \theta_N; t)$, which evolves according to the Fokker-Planck equation:
\begin{equation}
    \frac{\partial Q}{\partial t} =  
     \frac{1}{\pi_r}\frac{\partial}{\partial \theta_i}
    \left( \frac{\partial V}{\partial \theta_i} Q  
    + \pi_{\theta}  \frac{\partial Q}{\partial \theta_i}\right)-\boldsymbol{\nabla}_{\boldsymbol{x}_i} 
    \left(  \langle\boldsymbol{\varphi}^q | \boldsymbol{\hat{n}}\rangle  
    \boldsymbol{\varphi}_i^q Q\right) 
    \label{Fokker}
\end{equation}
with implicit summation on the indices.
The evolution of the $M$ zero modes can be described with a set of angles, distances, or more general coordinates, which we denote $\boldsymbol{\alpha} = \{\alpha_m\}_{m=1}^M$. 
To make further progress, we proceed to an adiabatic approximation, assuming that the dynamics of the zero modes is much slower than that of the orientation of the active units, which is expected to hold in the $\pi_r \ll 1$ regime.
At zeroth-order, this approximation amounts to considering that the probability density function $Q$ is different from zero only for combinations of $\boldsymbol{x}_i$ which preserve the same zero modes. 
In such a case, the Fokker-Planck equation for the reduced density probability function $\mathcal{Q}=\int_{-\infty}^{\infty}Q d\boldsymbol{x}_1\ldots d\boldsymbol{x}_N$, admits a steady state solution given by the Gibbs measure: 
\begin{equation} \label{eq:gibbs_solution}
\mathcal{Q} = \frac{\exp(-\beta V)}{\mathcal{Z}}
\end{equation}
with $\beta = 1/\pi_\theta$ and $\mathcal{Z} = \int_{-\pi}^{\pi}  e^{-\beta V}\ \mathrm{d} \theta_1 \dots \mathrm{d} \theta_N$. 

Collective motion is achieved when the normalized projections of the polarity vectors over the zero modes, namely the order parameters $\mu_q = \left \langle \left(\sum_i \boldsymbol{\varphi}^q_i \cdot \boldsymbol{\hat{n}}_i\right) \right \rangle / \sqrt{N}$ are $\mathcal{O}(1)$.   
In the thermodynamic limit, and considering the case of extended zero modes, such as translations, rotations, or auxetic modes~\cite{Acuna2022}, we find that the mode selection is governed by the minimum of the Landau free energy:
\begin{equation}
    f[\boldsymbol{\mu},\boldsymbol{\alpha}] =  \sum_{q \in \mathfrak{F}} \frac{\mu_q^2}{2}
    -
    \frac{1}{\beta N} \sum_i \log \left( I_0 \left( \beta 
    \mathcal{D}_i \right) \right),
\end{equation}
where $I_0$ is the modified Bessel function of the first kind and:
\begin{equation}
\mathcal{D}_i=\left(N\sum_{q,l \in \mathfrak{F}} \mu_q \mu_l \left(  \boldsymbol{\varphi}_i^q (\boldsymbol{\alpha}) \cdot \boldsymbol{\varphi}_i^l (\boldsymbol{\alpha}) \right)\right)^{1/2}
\end{equation}
couples the different zero modes. Finally, having found the order parameters for a particular system configuration $\boldsymbol{\alpha}$, 
we can 
evolve every $\alpha_m$ as $\dot{\alpha}_m = L_m (\boldsymbol{\alpha},\boldsymbol{\mu})$, where $L_m$ is a structure-dependent operator. 

Using the above formulation we can prove that, within the adiabatic approximation, for any system with extended zero modes, there is a continuous phase transition from a disordered phase to some form of collective motion, taking place at $\pi_{\theta}^c=1/2$. This description, valid only in the large $N$ limit, can be extended to small $N$ in terms of simple integrals 
and higher order corrections as powers of $\pi_r$ can also be found (See Sup. Mat. \cite{SM}, sections IV and VI). 

The simplest scenario corresponds to pure translational motion. If no other zero mode is allowed, as is the case for an unpinned lattice with PBC, the zeroth-order solution (Eq.~\ref{eq:gibbs_solution}) is exact because the translational modes are position independent. The order parameter is $\boldsymbol{\mu}_P = \left(\mu_x,\mu_y\right) = \left\langle \frac{1}{N} \sum_i \hat{\boldsymbol{n}}_i \right\rangle$. The angular potential reads $V= - \frac{1}{2N} \sum_{i,j} \cos(\theta_i - \theta_j)$ and therefore exactly maps onto the 2D mean-field XY model. The minima of the corresponding free energy shown in Fig. \ref{fig:fig2}-d for two noise amplitudes, perfectly captures the transition. In the thermodynamic limit, a phase transition at $\pi_{\theta}=1/2$ with the mean-field critical exponents is obtained (See Sup. Mat. \cite{SM}, section IV). 
Remarkably, the mean-field behavior does not arise from an uncontrolled approximation: true long-range order emerges from system-wide stress propagation, resulting from rigidity (See Sup. Movie 2 \cite{SM}).

In the purely rotational, or auxetic cases, 
the adiabatic approximation is not exact because the associated zero modes depend on the instantaneous structure prescribed by $\phi$, the rotational angle, and $\gamma$, the compression angle (See Fig. \ref{fig:fig2}-e and Fig. \ref{fig:fig2}-i). However, in the presence of a  single zero mode, an adequate choice of reference frame
removes this dependence.
Here also we find a perfect agreement between the simulation data (Fig~\ref{fig:fig2}-g,k), and the order parameters $\mu_{\phi}$ and $\mu_{\gamma}$ extracted from the minimization of the free energy shown on Fig~\ref{fig:fig2}-h,l 
(See Sup. Mat. \cite{SM}, sections VI and VII). The dependence of these so-defined order parameters on $\pi_r$, when $\pi_\theta=0$, is also perfectly 
captured (See Fig~\ref{fig:fig2}-g,k and Sup. Mat. \cite{SM}, sections VI and VII).  

Two-mode settings lead to more intricate dynamics and complex energy landscapes. First, we will consider a translational-rotational system that has a Landau free energy that is independent of the structure parameter,  an exception due to the structure-invariant nature of the translational modes. The order parameters are the ones defined for the pure translation and rotational case. The free energy, however, displays a richer behavior, when  $\pi_\theta < \pi_{\theta}^c = 1/2$ : the space isotropy for translation and the chiral symmetry for rotation are simultaneously broken (See Fig.~\ref{fig:fig3}-b and Sup. Movie 3).
Interestingly, the translational solution is always the global minimum (See Fig.~\ref{fig:fig3}-c and Sup. Mat. \cite{SM}, section VIII), and mixed translational/rotational states are not steady state solutions.
The mean-field nature of the system leads to minima that are separated by an energy barrier proportional to $N$. 
 
Finally, considering a network where the two zero modes are the rotation and the auxetic one allows us to demonstrate the efficiency of our approach, while stressing its limitations. 
Our prescription eliminates the dependence on $\phi$ and defines the two order parameters
$\mu_A$ and $\mu_B$ as functions of $\langle \dot{\phi} \rangle$, $\langle \dot{\gamma} \rangle$ and $\gamma$ (See Sup. Mat. \cite{SM}, section IX).
Depending on the value of $\pi_{\theta}$, the free energy has 4 local minima, 2, or just one, i.e. the disordered solution. 
As shown in Fig. \ref{fig:fig3}-f  these minima move in phase space as $\gamma$ evolves.  For a given set $\pi_{\theta},\pi_r$, we can follow the evolution of each different solution from the adiabatic prescription $\dot{\alpha}_m = L_m (\boldsymbol{\alpha},\boldsymbol{\mu})$ (See Sup. Mat. \cite{SM}, section IX).
This adiabatic evolution converges to a well-defined phase diagram (See Fig.~\ref{fig:fig3}-e) that remarkably displays two transitions, the aforementioned one at $\pi_\theta^c=1/2$, where the auxetic contraction is activated, and a second one at much smaller values, $\pi_\theta\approx 0.02$, where rotation is activated.  
Numerical simulations of the auxetic-rotation systems show good agreement in most of the phase diagram, except in the region below
the onset of rotational dynamics (See inset of Fig.~\ref{fig:fig3}-e), where the system evolves following two minima that are separated by a low, vanishing for large $N$, energy barrier. An appropriate numerical procedure to follow the evolution of the structure should solve this discrepancy.

In this letter, we have studied the dynamics of active solid body motion and mechanism folding through a general theoretical framework in the rigid limit. Our formalism allows for
the design and tuning of a wide range of materials where elastic deformations are negligible,
with the interaction between different modes giving rise to rich, complex dynamics. 
Future work will restore elasticity, in particular, elucidating whether long-range order is preserved for different system sizes and spring constants. Numerical studies on a related model seem to suggest that stiff enough systems display collective motion independently of the system size \cite{Couzin2023}, and in such a case, corrections beyond mean-field could be calculated.
Furthermore, we will explore structures with multiple shape-changing modes \cite{bossart_2021}, 
settings with various active solids or environmental obstacles,
and consider our dynamical system outside the $\pi_r \ll 1$ limit, potentially capturing 
other effects such as the geometry-controlled selection of particular translation directions~\cite{ferrante_elasticity-based_2013}.

C.H-L. was supported by a Ph.D. grant from ED564 ‘Physique en Ile de France’. G.D. acknowledges support from  Fondecyt Grant No. 1210656. P.B. was supported by a Ph.D. grant from ED564 ‘Physique en Ile de France’. C.C. acknowledges funding from the European Research Council under grant agreement 852587 and from the Netherlands Organisation for Scientific Research under grant agreement VIDI 2131313.

\nocite{}
\bibliographystyle{apsrev4-2}
\vspace{-5mm}
\bibliography{bibliography.bib, Active.bib}

\begin{thebibliography}{33}%
\makeatletter
\providecommand \@ifxundefined [1]{%
 \@ifx{#1\undefined}
}%
\providecommand \@ifnum [1]{%
 \ifnum #1\expandafter \@firstoftwo
 \else \expandafter \@secondoftwo
 \fi
}%
\providecommand \@ifx [1]{%
 \ifx #1\expandafter \@firstoftwo
 \else \expandafter \@secondoftwo
 \fi
}%
\providecommand \natexlab [1]{#1}%
\providecommand \enquote  [1]{``#1''}%
\providecommand \bibnamefont  [1]{#1}%
\providecommand \bibfnamefont [1]{#1}%
\providecommand \citenamefont [1]{#1}%
\providecommand \href@noop [0]{\@secondoftwo}%
\providecommand \href [0]{\begingroup \@sanitize@url \@href}%
\providecommand \@href[1]{\@@startlink{#1}\@@href}%
\providecommand \@@href[1]{\endgroup#1\@@endlink}%
\providecommand \@sanitize@url [0]{\catcode `\\12\catcode `\$12\catcode
  `\&12\catcode `\#12\catcode `\^12\catcode `\_12\catcode `\%12\relax}%
\providecommand \@@startlink[1]{}%
\providecommand \@@endlink[0]{}%
\providecommand \url  [0]{\begingroup\@sanitize@url \@url }%
\providecommand \@url [1]{\endgroup\@href {#1}{\urlprefix }}%
\providecommand \urlprefix  [0]{URL }%
\providecommand \Eprint [0]{\href }%
\providecommand \doibase [0]{https://doi.org/}%
\providecommand \selectlanguage [0]{\@gobble}%
\providecommand \bibinfo  [0]{\@secondoftwo}%
\providecommand \bibfield  [0]{\@secondoftwo}%
\providecommand \translation [1]{[#1]}%
\providecommand \BibitemOpen [0]{}%
\providecommand \bibitemStop [0]{}%
\providecommand \bibitemNoStop [0]{.\EOS\space}%
\providecommand \EOS [0]{\spacefactor3000\relax}%
\providecommand \BibitemShut  [1]{\csname bibitem#1\endcsname}%
\let\auto@bib@innerbib\@empty
\bibitem [{\citenamefont {Cavagna}\ \emph {et~al.}(2010)\citenamefont
  {Cavagna}, \citenamefont {Cimarelli}, \citenamefont {Giardina}, \citenamefont
  {Parisi}, \citenamefont {Santagati}, \citenamefont {Stefanini},\ and\
  \citenamefont {Viale}}]{Cavagna-2010}%
  \BibitemOpen
  \bibfield  {author} {\bibinfo {author} {\bibfnamefont {A.}~\bibnamefont
  {Cavagna}}, \bibinfo {author} {\bibfnamefont {A.}~\bibnamefont {Cimarelli}},
  \bibinfo {author} {\bibfnamefont {I.}~\bibnamefont {Giardina}}, \bibinfo
  {author} {\bibfnamefont {G.}~\bibnamefont {Parisi}}, \bibinfo {author}
  {\bibfnamefont {R.}~\bibnamefont {Santagati}}, \bibinfo {author}
  {\bibfnamefont {F.}~\bibnamefont {Stefanini}},\ and\ \bibinfo {author}
  {\bibfnamefont {M.}~\bibnamefont {Viale}},\ }\href
  {https://doi.org/10.1073/pnas.1005766107} {\bibfield  {journal} {\bibinfo
  {journal} {Proceedings of the National Academy of Sciences}\ }\textbf
  {\bibinfo {volume} {107}},\ \bibinfo {pages} {11865} (\bibinfo {year}
  {2010})}\BibitemShut {NoStop}%
\bibitem [{\citenamefont {Bialek}\ \emph {et~al.}(2012)\citenamefont {Bialek},
  \citenamefont {Cavagna}, \citenamefont {Giardina}, \citenamefont {Mora},
  \citenamefont {Silvestri}, \citenamefont {Viale},\ and\ \citenamefont
  {Walczak}}]{bialek_statistical_2012}%
  \BibitemOpen
  \bibfield  {author} {\bibinfo {author} {\bibfnamefont {W.}~\bibnamefont
  {Bialek}}, \bibinfo {author} {\bibfnamefont {A.}~\bibnamefont {Cavagna}},
  \bibinfo {author} {\bibfnamefont {I.}~\bibnamefont {Giardina}}, \bibinfo
  {author} {\bibfnamefont {T.}~\bibnamefont {Mora}}, \bibinfo {author}
  {\bibfnamefont {E.}~\bibnamefont {Silvestri}}, \bibinfo {author}
  {\bibfnamefont {M.}~\bibnamefont {Viale}},\ and\ \bibinfo {author}
  {\bibfnamefont {A.~M.}\ \bibnamefont {Walczak}},\ }\href
  {https://doi.org/10.1073/pnas.1118633109} {\bibfield  {journal} {\bibinfo
  {journal} {Proceedings of the National Academy of Sciences}\ }\textbf
  {\bibinfo {volume} {109}},\ \bibinfo {pages} {4786} (\bibinfo {year}
  {2012})}\BibitemShut {NoStop}%
\bibitem [{\citenamefont {Bricard}\ \emph {et~al.}(2013)\citenamefont
  {Bricard}, \citenamefont {Caussin}, \citenamefont {Desreumaux}, \citenamefont
  {Dauchot},\ and\ \citenamefont {Bartolo}}]{bricard_emergence_2013}%
  \BibitemOpen
  \bibfield  {author} {\bibinfo {author} {\bibfnamefont {A.}~\bibnamefont
  {Bricard}}, \bibinfo {author} {\bibfnamefont {J.-B.}\ \bibnamefont
  {Caussin}}, \bibinfo {author} {\bibfnamefont {N.}~\bibnamefont {Desreumaux}},
  \bibinfo {author} {\bibfnamefont {O.}~\bibnamefont {Dauchot}},\ and\ \bibinfo
  {author} {\bibfnamefont {D.}~\bibnamefont {Bartolo}},\ }\href
  {https://doi.org/10.1038/nature12673} {\bibfield  {journal} {\bibinfo
  {journal} {Nature}\ }\textbf {\bibinfo {volume} {503}},\ \bibinfo {pages}
  {95} (\bibinfo {year} {2013})}\BibitemShut {NoStop}%
\bibitem [{\citenamefont {Weber}\ \emph {et~al.}(2013)\citenamefont {Weber},
  \citenamefont {Hanke}, \citenamefont {Deseigne}, \citenamefont {Léonard},
  \citenamefont {Dauchot}, \citenamefont {Frey},\ and\ \citenamefont
  {Chaté}}]{weber_long-range_2013}%
  \BibitemOpen
  \bibfield  {author} {\bibinfo {author} {\bibfnamefont {C.~A.}\ \bibnamefont
  {Weber}}, \bibinfo {author} {\bibfnamefont {T.}~\bibnamefont {Hanke}},
  \bibinfo {author} {\bibfnamefont {J.}~\bibnamefont {Deseigne}}, \bibinfo
  {author} {\bibfnamefont {S.}~\bibnamefont {Léonard}}, \bibinfo {author}
  {\bibfnamefont {O.}~\bibnamefont {Dauchot}}, \bibinfo {author} {\bibfnamefont
  {E.}~\bibnamefont {Frey}},\ and\ \bibinfo {author} {\bibfnamefont
  {H.}~\bibnamefont {Chaté}},\ }\href
  {https://doi.org/10.1103/PhysRevLett.110.208001} {\bibfield  {journal}
  {\bibinfo  {journal} {Physical Review Letters}\ }\textbf {\bibinfo {volume}
  {110}},\ \bibinfo {pages} {208001} (\bibinfo {year} {2013})}\BibitemShut
  {NoStop}%
\bibitem [{\citenamefont {Ferrante}\ \emph
  {et~al.}(2013{\natexlab{a}})\citenamefont {Ferrante}, \citenamefont {Turgut},
  \citenamefont {Dorigo},\ and\ \citenamefont
  {Huepe}}]{ferrante_elasticity-based_2013}%
  \BibitemOpen
  \bibfield  {author} {\bibinfo {author} {\bibfnamefont {E.}~\bibnamefont
  {Ferrante}}, \bibinfo {author} {\bibfnamefont {A.~E.}\ \bibnamefont
  {Turgut}}, \bibinfo {author} {\bibfnamefont {M.}~\bibnamefont {Dorigo}},\
  and\ \bibinfo {author} {\bibfnamefont {C.}~\bibnamefont {Huepe}},\ }\href
  {https://doi.org/10.1103/PhysRevLett.111.268302} {\bibfield  {journal}
  {\bibinfo  {journal} {Phys. Rev. Lett.}\ }\textbf {\bibinfo {volume} {111}},\
  \bibinfo {pages} {268302} (\bibinfo {year} {2013}{\natexlab{a}})}\BibitemShut
  {NoStop}%
\bibitem [{\citenamefont {Zheng}\ \emph {et~al.}(2020)\citenamefont {Zheng},
  \citenamefont {Huepe},\ and\ \citenamefont {Han}}]{zheng_experimental_2020}%
  \BibitemOpen
  \bibfield  {author} {\bibinfo {author} {\bibfnamefont {Y.}~\bibnamefont
  {Zheng}}, \bibinfo {author} {\bibfnamefont {C.}~\bibnamefont {Huepe}},\ and\
  \bibinfo {author} {\bibfnamefont {Z.}~\bibnamefont {Han}},\ }\href
  {https://doi.org/10.1177/1059712320930418} {\bibfield  {journal} {\bibinfo
  {journal} {Adaptive Behavior}\ ,\ \bibinfo {pages} {1059712320930418}}
  (\bibinfo {year} {2020})}\BibitemShut {NoStop}%
\bibitem [{\citenamefont {Teixeira}\ \emph {et~al.}(2021)\citenamefont
  {Teixeira}, \citenamefont {Fernandes},\ and\ \citenamefont
  {Brunnet}}]{D1SM00080B}%
  \BibitemOpen
  \bibfield  {author} {\bibinfo {author} {\bibfnamefont {E.~F.}\ \bibnamefont
  {Teixeira}}, \bibinfo {author} {\bibfnamefont {H.~C.~M.}\ \bibnamefont
  {Fernandes}},\ and\ \bibinfo {author} {\bibfnamefont {L.~G.}\ \bibnamefont
  {Brunnet}},\ }\href {https://doi.org/10.1039/D1SM00080B} {\bibfield
  {journal} {\bibinfo  {journal} {Soft Matter}\ }\textbf {\bibinfo {volume}
  {17}},\ \bibinfo {pages} {5991} (\bibinfo {year} {2021})}\BibitemShut
  {NoStop}%
\bibitem [{\citenamefont {Flierl}\ \emph {et~al.}(1999)\citenamefont {Flierl},
  \citenamefont {Grünbaum}, \citenamefont {Levins},\ and\ \citenamefont
  {Olson}}]{flierl_individuals_1999}%
  \BibitemOpen
  \bibfield  {author} {\bibinfo {author} {\bibfnamefont {G.}~\bibnamefont
  {Flierl}}, \bibinfo {author} {\bibfnamefont {D.}~\bibnamefont {Grünbaum}},
  \bibinfo {author} {\bibfnamefont {S.}~\bibnamefont {Levins}},\ and\ \bibinfo
  {author} {\bibfnamefont {D.}~\bibnamefont {Olson}},\ }\href
  {https://doi.org/https://doi.org/10.1006/jtbi.1998.0842} {\bibfield
  {journal} {\bibinfo  {journal} {Journal of Theoretical Biology}\ }\textbf
  {\bibinfo {volume} {196}},\ \bibinfo {pages} {397 } (\bibinfo {year}
  {1999})}\BibitemShut {NoStop}%
\bibitem [{\citenamefont {Liu}\ \emph {et~al.}(2021)\citenamefont {Liu},
  \citenamefont {Shankar}, \citenamefont {Marchetti},\ and\ \citenamefont
  {Wu}}]{Liu-2021}%
  \BibitemOpen
  \bibfield  {author} {\bibinfo {author} {\bibfnamefont {S.}~\bibnamefont
  {Liu}}, \bibinfo {author} {\bibfnamefont {S.}~\bibnamefont {Shankar}},
  \bibinfo {author} {\bibfnamefont {M.~C.}\ \bibnamefont {Marchetti}},\ and\
  \bibinfo {author} {\bibfnamefont {Y.}~\bibnamefont {Wu}},\ }\href
  {https://doi.org/10.1038/s41586-020-03168-6} {\bibfield  {journal} {\bibinfo
  {journal} {Nature}\ }\textbf {\bibinfo {volume} {590}},\ \bibinfo {pages}
  {80} (\bibinfo {year} {2021})}\BibitemShut {NoStop}%
\bibitem [{\citenamefont {Bricard}\ \emph {et~al.}(2015)\citenamefont
  {Bricard}, \citenamefont {Caussin}, \citenamefont {Das}, \citenamefont
  {Savoie}, \citenamefont {Chikkadi}, \citenamefont {Shitara}, \citenamefont
  {Chepizhko}, \citenamefont {Peruani}, \citenamefont {Saintillan},\ and\
  \citenamefont {Bartolo}}]{Bricard-2015}%
  \BibitemOpen
  \bibfield  {author} {\bibinfo {author} {\bibfnamefont {A.}~\bibnamefont
  {Bricard}}, \bibinfo {author} {\bibfnamefont {J.-B.}\ \bibnamefont
  {Caussin}}, \bibinfo {author} {\bibfnamefont {D.}~\bibnamefont {Das}},
  \bibinfo {author} {\bibfnamefont {C.}~\bibnamefont {Savoie}}, \bibinfo
  {author} {\bibfnamefont {V.}~\bibnamefont {Chikkadi}}, \bibinfo {author}
  {\bibfnamefont {K.}~\bibnamefont {Shitara}}, \bibinfo {author} {\bibfnamefont
  {O.}~\bibnamefont {Chepizhko}}, \bibinfo {author} {\bibfnamefont
  {F.}~\bibnamefont {Peruani}}, \bibinfo {author} {\bibfnamefont
  {D.}~\bibnamefont {Saintillan}},\ and\ \bibinfo {author} {\bibfnamefont
  {D.}~\bibnamefont {Bartolo}},\ }\href {https://doi.org/10.1038/ncomms8470}
  {\bibfield  {journal} {\bibinfo  {journal} {Nature Communications}\ }\textbf
  {\bibinfo {volume} {6}},\ \bibinfo {pages} {1} (\bibinfo {year}
  {2015})}\BibitemShut {NoStop}%
\bibitem [{\citenamefont {Chardac}\ \emph {et~al.}(2021)\citenamefont
  {Chardac}, \citenamefont {Hoffmann}, \citenamefont {Poupart}, \citenamefont
  {Giomi},\ and\ \citenamefont {Bartolo}}]{Chardac-2021}%
  \BibitemOpen
  \bibfield  {author} {\bibinfo {author} {\bibfnamefont {A.}~\bibnamefont
  {Chardac}}, \bibinfo {author} {\bibfnamefont {L.~A.}\ \bibnamefont
  {Hoffmann}}, \bibinfo {author} {\bibfnamefont {Y.}~\bibnamefont {Poupart}},
  \bibinfo {author} {\bibfnamefont {L.}~\bibnamefont {Giomi}},\ and\ \bibinfo
  {author} {\bibfnamefont {D.}~\bibnamefont {Bartolo}},\ }\href
  {https://doi.org/10.1103/PhysRevX.11.031069} {\bibfield  {journal} {\bibinfo
  {journal} {Physical Review X}\ }\textbf {\bibinfo {volume} {11}},\ \bibinfo
  {pages} {031069} (\bibinfo {year} {2021})}\BibitemShut {NoStop}%
\bibitem [{\citenamefont {Toner}\ \emph {et~al.}(2005)\citenamefont {Toner},
  \citenamefont {Tu},\ and\ \citenamefont {Ramaswamy}}]{Toner-2005}%
  \BibitemOpen
  \bibfield  {author} {\bibinfo {author} {\bibfnamefont {J.}~\bibnamefont
  {Toner}}, \bibinfo {author} {\bibfnamefont {Y.}~\bibnamefont {Tu}},\ and\
  \bibinfo {author} {\bibfnamefont {S.}~\bibnamefont {Ramaswamy}},\ }\href
  {https://doi.org/10.1016/j.aop.2005.04.011} {\bibfield  {journal} {\bibinfo
  {journal} {Annals of Physics}\ }\textbf {\bibinfo {volume} {318}},\ \bibinfo
  {pages} {170} (\bibinfo {year} {2005})}\BibitemShut {NoStop}%
\bibitem [{\citenamefont {Toner}(2012)}]{Toner-2012}%
  \BibitemOpen
  \bibfield  {author} {\bibinfo {author} {\bibfnamefont {J.}~\bibnamefont
  {Toner}},\ }\href {https://doi.org/10.1103/PhysRevE.86.031918} {\bibfield
  {journal} {\bibinfo  {journal} {Physical Review E}\ }\textbf {\bibinfo
  {volume} {86}},\ \bibinfo {pages} {031918} (\bibinfo {year}
  {2012})}\BibitemShut {NoStop}%
\bibitem [{\citenamefont {Henkes}\ \emph {et~al.}(2011)\citenamefont {Henkes},
  \citenamefont {Fily},\ and\ \citenamefont {Marchetti}}]{henkes_active_2011}%
  \BibitemOpen
  \bibfield  {author} {\bibinfo {author} {\bibfnamefont {S.}~\bibnamefont
  {Henkes}}, \bibinfo {author} {\bibfnamefont {Y.}~\bibnamefont {Fily}},\ and\
  \bibinfo {author} {\bibfnamefont {M.~C.}\ \bibnamefont {Marchetti}},\ }\href
  {https://doi.org/10.1103/PhysRevE.84.040301} {\bibfield  {journal} {\bibinfo
  {journal} {Physical Review E}\ }\textbf {\bibinfo {volume} {84}},\ \bibinfo
  {pages} {040301} (\bibinfo {year} {2011})}\BibitemShut {NoStop}%
\bibitem [{\citenamefont {Shimoyama}\ \emph {et~al.}(1996)\citenamefont
  {Shimoyama}, \citenamefont {Sugawara}, \citenamefont {Mizuguchi},
  \citenamefont {Hayakawa},\ and\ \citenamefont
  {Sano}}]{shimoyama_collective_1996}%
  \BibitemOpen
  \bibfield  {author} {\bibinfo {author} {\bibfnamefont {N.}~\bibnamefont
  {Shimoyama}}, \bibinfo {author} {\bibfnamefont {K.}~\bibnamefont {Sugawara}},
  \bibinfo {author} {\bibfnamefont {T.}~\bibnamefont {Mizuguchi}}, \bibinfo
  {author} {\bibfnamefont {Y.}~\bibnamefont {Hayakawa}},\ and\ \bibinfo
  {author} {\bibfnamefont {M.}~\bibnamefont {Sano}},\ }\href
  {https://doi.org/10.1103/PhysRevLett.76.3870} {\bibfield  {journal} {\bibinfo
   {journal} {Physical Review Letters}\ }\textbf {\bibinfo {volume} {76}},\
  \bibinfo {pages} {3870} (\bibinfo {year} {1996})}\BibitemShut {NoStop}%
\bibitem [{\citenamefont {Grégoire}\ and\ \citenamefont
  {Chaté}(2004)}]{Gregoire-2004}%
  \BibitemOpen
  \bibfield  {author} {\bibinfo {author} {\bibfnamefont {G.}~\bibnamefont
  {Grégoire}}\ and\ \bibinfo {author} {\bibfnamefont {H.}~\bibnamefont
  {Chaté}},\ }\href {https://doi.org/10.1103/PhysRevLett.92.025702} {\bibfield
   {journal} {\bibinfo  {journal} {Physical Review Letters}\ }\textbf {\bibinfo
  {volume} {92}},\ \bibinfo {pages} {025702} (\bibinfo {year}
  {2004})}\BibitemShut {NoStop}%
\bibitem [{\citenamefont {Xu}\ \emph {et~al.}(2023)\citenamefont {Xu},
  \citenamefont {Huang}, \citenamefont {Zhang},\ and\ \citenamefont
  {Wu}}]{Xu-2023}%
  \BibitemOpen
  \bibfield  {author} {\bibinfo {author} {\bibfnamefont {H.}~\bibnamefont
  {Xu}}, \bibinfo {author} {\bibfnamefont {Y.}~\bibnamefont {Huang}}, \bibinfo
  {author} {\bibfnamefont {R.}~\bibnamefont {Zhang}},\ and\ \bibinfo {author}
  {\bibfnamefont {Y.}~\bibnamefont {Wu}},\ }\href
  {https://doi.org/10.1038/s41567-022-01836-0} {\bibfield  {journal} {\bibinfo
  {journal} {Nature Physics}\ }\textbf {\bibinfo {volume} {19}},\ \bibinfo
  {pages} {46} (\bibinfo {year} {2023})}\BibitemShut {NoStop}%
\bibitem [{\citenamefont {Szabó}\ \emph {et~al.}(2006)\citenamefont {Szabó},
  \citenamefont {Szöllösi}, \citenamefont {Gönci}, \citenamefont {Jurányi},
  \citenamefont {Selmeczi},\ and\ \citenamefont {Vicsek}}]{szabo_phase_2006}%
  \BibitemOpen
  \bibfield  {author} {\bibinfo {author} {\bibfnamefont {B.}~\bibnamefont
  {Szabó}}, \bibinfo {author} {\bibfnamefont {G.~J.}\ \bibnamefont
  {Szöllösi}}, \bibinfo {author} {\bibfnamefont {B.}~\bibnamefont {Gönci}},
  \bibinfo {author} {\bibfnamefont {Z.}~\bibnamefont {Jurányi}}, \bibinfo
  {author} {\bibfnamefont {D.}~\bibnamefont {Selmeczi}},\ and\ \bibinfo
  {author} {\bibfnamefont {T.}~\bibnamefont {Vicsek}},\ }\href
  {https://doi.org/10.1103/PhysRevE.74.061908} {\bibfield  {journal} {\bibinfo
  {journal} {Physical Review E}\ }\textbf {\bibinfo {volume} {74}},\ \bibinfo
  {pages} {061908} (\bibinfo {year} {2006})}\BibitemShut {NoStop}%
\bibitem [{\citenamefont {Peyret}\ \emph {et~al.}(2019)\citenamefont {Peyret},
  \citenamefont {Mueller}, \citenamefont {d’Alessandro}, \citenamefont
  {Begnaud}, \citenamefont {Marcq}, \citenamefont {Mège}, \citenamefont
  {Yeomans}, \citenamefont {Doostmohammadi},\ and\ \citenamefont
  {Ladoux}}]{peyret_sustained_2019}%
  \BibitemOpen
  \bibfield  {author} {\bibinfo {author} {\bibfnamefont {G.}~\bibnamefont
  {Peyret}}, \bibinfo {author} {\bibfnamefont {R.}~\bibnamefont {Mueller}},
  \bibinfo {author} {\bibfnamefont {J.}~\bibnamefont {d’Alessandro}},
  \bibinfo {author} {\bibfnamefont {S.}~\bibnamefont {Begnaud}}, \bibinfo
  {author} {\bibfnamefont {P.}~\bibnamefont {Marcq}}, \bibinfo {author}
  {\bibfnamefont {R.-M.}\ \bibnamefont {Mège}}, \bibinfo {author}
  {\bibfnamefont {J.~M.}\ \bibnamefont {Yeomans}}, \bibinfo {author}
  {\bibfnamefont {A.}~\bibnamefont {Doostmohammadi}},\ and\ \bibinfo {author}
  {\bibfnamefont {B.}~\bibnamefont {Ladoux}},\ }\href
  {https://doi.org/10.1016/j.bpj.2019.06.013} {\bibfield  {journal} {\bibinfo
  {journal} {Biophysical Journal}\ }\textbf {\bibinfo {volume} {117}},\
  \bibinfo {pages} {464} (\bibinfo {year} {2019})}\BibitemShut {NoStop}%
\bibitem [{\citenamefont {Ferrante}\ \emph
  {et~al.}(2013{\natexlab{b}})\citenamefont {Ferrante}, \citenamefont {Turgut},
  \citenamefont {Dorigo},\ and\ \citenamefont {Huepe}}]{Ferrante-2013a}%
  \BibitemOpen
  \bibfield  {author} {\bibinfo {author} {\bibfnamefont {E.}~\bibnamefont
  {Ferrante}}, \bibinfo {author} {\bibfnamefont {A.~E.}\ \bibnamefont
  {Turgut}}, \bibinfo {author} {\bibfnamefont {M.}~\bibnamefont {Dorigo}},\
  and\ \bibinfo {author} {\bibfnamefont {C.}~\bibnamefont {Huepe}},\ }\href
  {https://doi.org/10.1103/PhysRevLett.111.268302} {\bibfield  {journal}
  {\bibinfo  {journal} {Physical Review Letters}\ }\textbf {\bibinfo {volume}
  {111}},\ \bibinfo {pages} {268302} (\bibinfo {year}
  {2013}{\natexlab{b}})}\BibitemShut {NoStop}%
\bibitem [{\citenamefont {Woodhouse}\ \emph {et~al.}(2018)\citenamefont
  {Woodhouse}, \citenamefont {Ronellenfitsch},\ and\ \citenamefont
  {Dunkel}}]{Woodhouse-2018}%
  \BibitemOpen
  \bibfield  {author} {\bibinfo {author} {\bibfnamefont {F.~G.}\ \bibnamefont
  {Woodhouse}}, \bibinfo {author} {\bibfnamefont {H.}~\bibnamefont
  {Ronellenfitsch}},\ and\ \bibinfo {author} {\bibfnamefont {J.}~\bibnamefont
  {Dunkel}},\ }\href {https://doi.org/10.1103/PhysRevLett.121.178001}
  {\bibfield  {journal} {\bibinfo  {journal} {Physical Review Letters}\
  }\textbf {\bibinfo {volume} {121}},\ \bibinfo {pages} {178001} (\bibinfo
  {year} {2018})}\BibitemShut {NoStop}%
\bibitem [{\citenamefont {Baconnier}\ \emph {et~al.}(2022)\citenamefont
  {Baconnier}, \citenamefont {Shohat}, \citenamefont {López}, \citenamefont
  {Coulais}, \citenamefont {Démery}, \citenamefont {Düring},\ and\
  \citenamefont {Dauchot}}]{baconnier_selective_2022}%
  \BibitemOpen
  \bibfield  {author} {\bibinfo {author} {\bibfnamefont {P.}~\bibnamefont
  {Baconnier}}, \bibinfo {author} {\bibfnamefont {D.}~\bibnamefont {Shohat}},
  \bibinfo {author} {\bibfnamefont {C.~H.}\ \bibnamefont {López}}, \bibinfo
  {author} {\bibfnamefont {C.}~\bibnamefont {Coulais}}, \bibinfo {author}
  {\bibfnamefont {V.}~\bibnamefont {Démery}}, \bibinfo {author} {\bibfnamefont
  {G.}~\bibnamefont {Düring}},\ and\ \bibinfo {author} {\bibfnamefont
  {O.}~\bibnamefont {Dauchot}},\ }\href
  {https://doi.org/10.1038/s41567-022-01704-x} {\bibfield  {journal} {\bibinfo
  {journal} {Nature Physics}\ }\textbf {\bibinfo {volume} {18}},\ \bibinfo
  {pages} {1234} (\bibinfo {year} {2022})}\BibitemShut {NoStop}%
\bibitem [{\citenamefont {Zheng}\ \emph {et~al.}(2023)\citenamefont {Zheng},
  \citenamefont {Brandenbourger}, \citenamefont {Robinet}, \citenamefont
  {Schall}, \citenamefont {Lerner},\ and\ \citenamefont
  {Coulais}}]{Zheng-2023}%
  \BibitemOpen
  \bibfield  {author} {\bibinfo {author} {\bibfnamefont {E.}~\bibnamefont
  {Zheng}}, \bibinfo {author} {\bibfnamefont {M.}~\bibnamefont
  {Brandenbourger}}, \bibinfo {author} {\bibfnamefont {L.}~\bibnamefont
  {Robinet}}, \bibinfo {author} {\bibfnamefont {P.}~\bibnamefont {Schall}},
  \bibinfo {author} {\bibfnamefont {E.}~\bibnamefont {Lerner}},\ and\ \bibinfo
  {author} {\bibfnamefont {C.}~\bibnamefont {Coulais}},\ }\href
  {https://doi.org/10.1103/PhysRevLett.130.178202} {\bibfield  {journal}
  {\bibinfo  {journal} {Physical Review Letters}\ }\textbf {\bibinfo {volume}
  {130}},\ \bibinfo {pages} {178202} (\bibinfo {year} {2023})}\BibitemShut
  {NoStop}%
\bibitem [{\citenamefont {Mao}\ and\ \citenamefont
  {Lubensky}(2018)}]{Xiaoming2018}%
  \BibitemOpen
  \bibfield  {author} {\bibinfo {author} {\bibfnamefont {X.}~\bibnamefont
  {Mao}}\ and\ \bibinfo {author} {\bibfnamefont {T.~C.}\ \bibnamefont
  {Lubensky}},\ }\href
  {https://doi.org/10.1146/annurev-conmatphys-033117-054235} {\bibfield
  {journal} {\bibinfo  {journal} {Annual Review of Condensed Matter Physics}\
  }\textbf {\bibinfo {volume} {9}},\ \bibinfo {pages} {413} (\bibinfo {year}
  {2018})}\BibitemShut {NoStop}%
\bibitem [{\citenamefont {Davidescu}\ \emph {et~al.}(2023)\citenamefont
  {Davidescu}, \citenamefont {Romanczuk}, \citenamefont {Gregor},\ and\
  \citenamefont {Couzin}}]{Couzin2023}%
  \BibitemOpen
  \bibfield  {author} {\bibinfo {author} {\bibfnamefont {M.}~\bibnamefont
  {Davidescu}}, \bibinfo {author} {\bibfnamefont {P.}~\bibnamefont
  {Romanczuk}}, \bibinfo {author} {\bibfnamefont {T.}~\bibnamefont {Gregor}},\
  and\ \bibinfo {author} {\bibfnamefont {I.}~\bibnamefont {Couzin}},\ }\href
  {https://doi.org/10.1073/pnas.2206163120} {\bibfield  {journal} {\bibinfo
  {journal} {Proceedings of the National Academy of Sciences}\ }\textbf
  {\bibinfo {volume} {120}},\ \bibinfo {pages} {e2206163120} (\bibinfo {year}
  {2023})}\BibitemShut {NoStop}%
\bibitem [{SM()}]{SM}%
  \BibitemOpen
  \href@noop {} {}\bibinfo {note} {See Supplemental Material for mathematical
  derivations and movies}\BibitemShut {NoStop}%
\bibitem [{\citenamefont {Resch}(1965)}]{resch1965geometrical}%
  \BibitemOpen
  \bibfield  {author} {\bibinfo {author} {\bibfnamefont {R.~D.}\ \bibnamefont
  {Resch}},\ }\href
  {https://image-ppubs.uspto.gov/dirsearch-public/print/downloadPdf/3201894}
  {\bibinfo {title} {Geometrical device having articulated relatively movable
  sections}} (\bibinfo {year} {1965}),\ \bibinfo {note} {{U}nited States of
  America Patent 3201894 (1965)}\BibitemShut {NoStop}%
\bibitem [{\citenamefont {Grima}\ and\ \citenamefont
  {Evans}(2000)}]{grima2000auxetic}%
  \BibitemOpen
  \bibfield  {author} {\bibinfo {author} {\bibfnamefont {J.~N.}\ \bibnamefont
  {Grima}}\ and\ \bibinfo {author} {\bibfnamefont {K.~E.}\ \bibnamefont
  {Evans}},\ }\href {https://doi.org/10.1023/A:1006781224002} {\bibfield
  {journal} {\bibinfo  {journal} {Journal of Materials Science Letters}\
  }\textbf {\bibinfo {volume} {19}},\ \bibinfo {pages} {1563} (\bibinfo {year}
  {2000})}\BibitemShut {NoStop}%
\bibitem [{\citenamefont {Acuna}\ \emph {et~al.}(2022)\citenamefont {Acuna},
  \citenamefont {Guti{\'e}rrez}, \citenamefont {Silva}, \citenamefont {Palza},
  \citenamefont {Nunez},\ and\ \citenamefont {D{\"u}ring}}]{Acuna2022}%
  \BibitemOpen
  \bibfield  {author} {\bibinfo {author} {\bibfnamefont {D.}~\bibnamefont
  {Acuna}}, \bibinfo {author} {\bibfnamefont {F.}~\bibnamefont
  {Guti{\'e}rrez}}, \bibinfo {author} {\bibfnamefont {R.}~\bibnamefont
  {Silva}}, \bibinfo {author} {\bibfnamefont {H.}~\bibnamefont {Palza}},
  \bibinfo {author} {\bibfnamefont {A.~S.}\ \bibnamefont {Nunez}},\ and\
  \bibinfo {author} {\bibfnamefont {G.}~\bibnamefont {D{\"u}ring}},\ }\href
  {https://doi.org/10.1038/s42005-022-00876-5} {\bibfield  {journal} {\bibinfo
  {journal} {Communications Physics}\ }\textbf {\bibinfo {volume} {5}},\
  \bibinfo {pages} {1} (\bibinfo {year} {2022})}\BibitemShut {NoStop}%
\bibitem [{\citenamefont {Czajkowski}\ \emph {et~al.}(2022)\citenamefont
  {Czajkowski}, \citenamefont {Coulais}, \citenamefont {van Hecke},\ and\
  \citenamefont {Rocklin}}]{Czajkowski-2022}%
  \BibitemOpen
  \bibfield  {author} {\bibinfo {author} {\bibfnamefont {M.}~\bibnamefont
  {Czajkowski}}, \bibinfo {author} {\bibfnamefont {C.}~\bibnamefont {Coulais}},
  \bibinfo {author} {\bibfnamefont {M.}~\bibnamefont {van Hecke}},\ and\
  \bibinfo {author} {\bibfnamefont {D.}~\bibnamefont {Rocklin}},\ }\href
  {https://doi.org/10.1038/s41467-021-27825-0} {\bibfield  {journal} {\bibinfo
  {journal} {Nature Communications}\ }\textbf {\bibinfo {volume} {13}},\
  \bibinfo {pages} {1} (\bibinfo {year} {2022})}\BibitemShut {NoStop}%
\bibitem [{\citenamefont {Brandenbourger}\ \emph {et~al.}(2021)\citenamefont
  {Brandenbourger}, \citenamefont {Scheibner}, \citenamefont {Veenstra},
  \citenamefont {Vitelli},\ and\ \citenamefont
  {Coulais}}]{brandenbourger_limit_2021}%
  \BibitemOpen
  \bibfield  {author} {\bibinfo {author} {\bibfnamefont {M.}~\bibnamefont
  {Brandenbourger}}, \bibinfo {author} {\bibfnamefont {C.}~\bibnamefont
  {Scheibner}}, \bibinfo {author} {\bibfnamefont {J.}~\bibnamefont {Veenstra}},
  \bibinfo {author} {\bibfnamefont {V.}~\bibnamefont {Vitelli}},\ and\ \bibinfo
  {author} {\bibfnamefont {C.}~\bibnamefont {Coulais}},\ }\href
  {https://doi.org/10.48550/arXiv.2108.08837} {\bibfield  {journal} {\bibinfo
  {journal} {arXiv preprint arXiv:2108.08837}\ } (\bibinfo {year}
  {2021})}\BibitemShut {NoStop}%
\bibitem [{\citenamefont {Dauchot}\ and\ \citenamefont
  {Démery}(2019)}]{dauchot_dynamics_2019}%
  \BibitemOpen
  \bibfield  {author} {\bibinfo {author} {\bibfnamefont {O.}~\bibnamefont
  {Dauchot}}\ and\ \bibinfo {author} {\bibfnamefont {V.}~\bibnamefont
  {Démery}},\ }\href {https://doi.org/10.1103/PhysRevLett.122.068002}
  {\bibfield  {journal} {\bibinfo  {journal} {Physical Review Letters}\
  }\textbf {\bibinfo {volume} {122}},\ \bibinfo {pages} {068002} (\bibinfo
  {year} {2019})}\BibitemShut {NoStop}%
\bibitem [{\citenamefont {Bossart}\ \emph {et~al.}(2021)\citenamefont
  {Bossart}, \citenamefont {Dykstra}, \citenamefont {van~der Laan},\ and\
  \citenamefont {Coulais}}]{bossart_2021}%
  \BibitemOpen
  \bibfield  {author} {\bibinfo {author} {\bibfnamefont {A.}~\bibnamefont
  {Bossart}}, \bibinfo {author} {\bibfnamefont {D.~M.~J.}\ \bibnamefont
  {Dykstra}}, \bibinfo {author} {\bibfnamefont {J.}~\bibnamefont {van~der
  Laan}},\ and\ \bibinfo {author} {\bibfnamefont {C.}~\bibnamefont {Coulais}},\
  }\href {https://doi.org/10.1073/pnas.2018610118} {\bibfield  {journal}
  {\bibinfo  {journal} {Proceedings of the National Academy of Sciences}\
  }\textbf {\bibinfo {volume} {118}},\ \bibinfo {pages} {e2018610118} (\bibinfo
  {year} {2021})}\BibitemShut {NoStop}%
\end{thebibliography}%

\end{document}


\title{Active Solids Model: Rigid Body Motion and Shape-changing Mechanisms \\ Supplementary Material}
\author{Claudio Hern\'andez-L\'opez$^{1,4}$ }
\author{Paul Baconnier$^2$ }
\author{Corentin Coulais$^3$}
\author{Olivier Dauchot$^2$ }
\author{Gustavo Düring$^4$}
\affiliation {$^1$ Laboratoire de Physique de l’\'Ecole Normale Sup\'erieure,  UMR CNRS 8023, Université PSL, Sorbonne Université, 75005 Paris, France}
\affiliation{$^2$ Gulliver UMR CNRS 7083, ESPCI Paris, Universit\'e PSL, Paris, France}
\affiliation{$^3$Institute of Physics, Universiteit van Amsterdam, 1098 XH Amsterdam, The Netherlands}
\affiliation{$^4$Instituto de Física, Pontificia Universidad Católica de Chile, 8331150 Santiago, Chile}

\maketitle

\section{Overdamped N-particle dynamics}
\subsection*{Non-dimensionalization}
The dimensional form of the overdamped dynamics can be written as:
\begin{align}
\Gamma \dot{\boldsymbol{x} _i} &=  F_0 \boldsymbol{\hat{n}}_i + \boldsymbol{F}_i \label{eq:langevin_x} \\
\dot{\boldsymbol{\hat{n}}}_i &= \frac{1}{\ell_a}\left(\boldsymbol{\hat{n}}_i \times \dot{\boldsymbol{x}}_i\right) \times \boldsymbol{\hat{n}}_i + \sqrt{2 D_{\theta}} \xi_i \boldsymbol{\hat{n}}_i^{\perp} \label{eq:langevin_n}
\end{align} 
with $\boldsymbol{x}_i$, respectively $\boldsymbol{\hat{n}}_i = (\cos\theta_i, \sin\theta_i)$, the position and the polarization unit vector of active unit $i$,  $F_0$ the active force, $\Gamma$ a friction coefficient, and $\boldsymbol{F}_i$ the net external force
on unit $i$. 
The alignment length $\ell_a$ is the distance required to align the polarization towards the velocity due to the torque term, 
and $D_{\theta}$ is the angular diffusion coefficient, with $\xi_i$ a delta-correlated Gaussian white noise process.
Noting that $\dot{\boldsymbol{\hat{n}}}_i = \dot{\theta}_i \boldsymbol{\hat{n}}_i^{\perp}$, we can write:
\begin{equation}
\dot{\theta}_i = \frac{1}{\ell_a} (\boldsymbol{\hat{n}^\perp}_i \cdot \boldsymbol{F}_i)
+ \sqrt{2 D_{\theta}} \xi_i \label{eq:langevin_theta}
\end{equation}

The equations are made dimensionless as follows: $\boldsymbol{x}\rightarrow \ell_0 \boldsymbol{x}$, $\boldsymbol{F}\rightarrow F_0 \boldsymbol{F}$ and $t\rightarrow \frac{\ell_0}{v_0} t$,  where  $v_0=F_0/\Gamma$ is the velocity of a free active unit, and $\ell_0$ is the characteristic particle-particle distance. Under such rescaling, Eqs. (\ref{eq:langevin_x}, \ref{eq:langevin_theta})  read
\begin{align}
	\dot{\boldsymbol{x}}_i &=  \boldsymbol{\hat{n}}_i + \boldsymbol{F}_i, \label{eq:dimensionless_x} \\
	 \dot{\theta}_i &= \frac{1}{\pi_r} (\boldsymbol{\hat{n}^\perp}_i \cdot \boldsymbol{F}_i)
  + \sqrt{2 \pi_{\theta} / \pi_r} \xi_i, \label{eq:dimensionless_n}
\end{align}
where $\pi_r = \ell_a/\ell_0$ and $\pi_{\theta} = D_{\theta} \ell_a/v_0$.

\subsection*{Rigid limit in an elastic system}

We will prove that if the parameter $\pi = \ell_e/\ell_a$ is small compared to the smallest non-zero normal frequency
of the structure, then no elastic mode is excited due to activity, i.e. the behavior of the system is rigid. 
Assuming that every spring has the same rest length $\ell_0$, the interaction between two
agents, in physical units, can be represented as:
\begin{equation}
    f_{ij} = -k \left(\lVert \boldsymbol{x}_{ij} \rVert - \ell_0 \right),
\end{equation}
where $k$ is the spring constant and 
$\boldsymbol{x}_{ij} = \boldsymbol{x}_{i} - \boldsymbol{x}_{j}$. Hence the interaction force
on $i$ can be written as:
\begin{align}
    \boldsymbol{F}_{i} &= 
    \sum_{j \in V_i} f_{ij} \frac{\boldsymbol{x}_{ij}}{\lVert \boldsymbol{x}_{ij} \rVert}, \\
    &= \label{eq:elastic_noTaylor}
    - k \sum_{j \in V_i} \left(\lVert \boldsymbol{x}_{ij} \rVert - \ell_0 \right) \frac{\boldsymbol{x}_{ij}}{\lVert \boldsymbol{x}_{ij} \rVert},
\end{align}
where $V_i$ contains all the neighbors of $i$.

Let $\boldsymbol{u}_{i} = \boldsymbol{x}_{i} - \boldsymbol{x}_{i}^0$ be the displacement of particle $i$ with respect to the reference configuration. 
If we perform a Taylor expansion of Eq. \ref{eq:elastic_noTaylor} to study our system in the linear elasticity regime:
\begin{equation} \label{eq:elastic_expansion_deltas}
    \boldsymbol{F}_{i} = -k \sum_{j \in V_i} \left[\frac{(\Delta x^0_{ij})^2}{\ell_0^2} \left(u^x_i - u^x_j \right) 
    + \frac{\Delta x^0_{ij} \Delta y^0_{ij}}{\ell_0^2} \left( u^y_i - u^y_j \right) \right] \boldsymbol{\hat{x}}
    - k \sum_{j \in V_i} \left[\frac{(\Delta y^0_{ij})^2}{\ell_0^2} \left(u^y_i - u^y_j \right) 
    + \frac{\Delta x^0_{ij} \Delta y^0_{ij}}{\ell_0^2} \left( u^x_i - u^x_j \right) \right] \boldsymbol{\hat{y}},
\end{equation}
where the parameters $\Delta x^0_{ij} = x^0_i - x^0_j$ and $\Delta y^0_{ij} = y^0_i - y^0_j$ define the geometry of the structure. These will define the elements of the dynamical matrix $\mathcal{M}_{ij}$:
\begin{equation}
    \boldsymbol{F}_{i} = -\sum_{j} \mathcal{M}_{ij} \boldsymbol{u}_{j}.
\end{equation}
After rescaling:
\begin{equation}
    \boldsymbol{F}_{i} = -\frac{\ell_0}{F_0} \sum_{j} \mathcal{M}_{ij} \boldsymbol{u}_{j}.
\end{equation}
Note that, for a single particle in 1D, $\ell_e = F_0/k$ is the characteristic distance that it will travel before
stopping due to the extension of the spring. Thus, 
the dimensionless parameter $\pi = \ell_e/\ell_a$ compares the characteristic
distances over which the system deforms and the distances over which the polarity vectors reorient. 
Together with the previously introduced parameter $\pi_r = \ell_a / \ell_0$, we may write:
\begin{equation} \label{eq:elastic_force}
    \boldsymbol{F}_{i} = -\frac{1}{\pi \pi_r k} \sum_{j} \mathcal{M}_{ij} \boldsymbol{u}_{j}.
\end{equation}
We will introduce bra-kets to simplify notation, so that we may refer to the displacement vector and polarity angle of 
each particle as:
\begin{align}
	\boldsymbol{u}_{i} &= \Braket{i | \boldsymbol{u}}, \\
	\theta_i &= \Braket{i | \theta}.
\end{align}

Then, the equations of motion can be recast as:
\begin{align}
	\label{ket_u_dot_before}
	\ket{\dot{\boldsymbol{u}}} &= \ket{\boldsymbol{\hat{n}}} + \ket{\boldsymbol{F}}, \\
	\label{ket_theta_dot_before}
	\ket{\dot{\theta}} &= \frac{1}{\pi_r} \mathcal{K} \ket{\boldsymbol{F}},
\end{align}
where $\mathcal{K}$ is such that:
\begin{equation}
	\Braket{i | \mathcal{K} | j} = \left(\boldsymbol{\hat{n}}_i^{\perp}\right)^T \delta_{ij},
\end{equation}
and using Eq. \ref{eq:elastic_force}:
\begin{align}
	\label{ket_u_dot}
	\ket{\dot{\boldsymbol{u}}} &= \ket{\boldsymbol{\hat{n}}} - \frac{1}{\pi \pi_r k}\mathcal{M} \ket{\boldsymbol{u}}, \\
	\label{ket_theta_dot}
	\ket{\dot{\theta}} &= - \frac{1}{\pi \pi_r^2 k} \mathcal{K} \mathcal{M} \ket{\boldsymbol{u}}.
\end{align}

We can decompose the displacement ket $\ket{\boldsymbol{u}}$ in the eigenkets of the elastic matrix $\mathcal{M}$:
\begin{equation}
	\ket{\boldsymbol{u}} = \sum_q a^u_q \ket{\boldsymbol{\varphi}^q}.
\end{equation}

We have the following dynamical equation for the coefficients $a^u_q (t)$:
\begin{equation} \label{dynamical_equation_auk}
	\dot{a^u_q} + \frac{1}{\pi \pi_r k} \omega^2_q a^u_q = \sum_i (\boldsymbol{\varphi}^q_i \cdot \boldsymbol{\hat{n}}_i),
\end{equation}
where $\omega_q^2$ is the $q$-th eigenvalue of $\mathcal{M}$, and the polarity angles $\theta_i (t)$:
\begin{equation} \label{dynamical_equation_thetai}
	\pi_r \dot{\theta}_i = - \frac{1}{\pi \pi_r k} \sum_q \omega_q^2 (\boldsymbol{\varphi}^q_i \cdot \boldsymbol{\hat{n}}_i^{\perp}) a_q^u.
\end{equation}

As explained in the main text, and from Eq. \ref{dynamical_equation_thetai}, the polarization angles evolve over a time
which scales as $\tilde{t} = t/\pi_r$. As such, we will rewrite Eq. \ref{dynamical_equation_auk} using the same timescale to analyze the dynamics:
\begin{equation}
    \frac{1}{\pi_r} \frac{\mathrm{d}}{\mathrm{d} \tilde{t}} a^u_q + 
    \frac{1}{\pi \pi_r k} \omega^2_q a^u_q = \sum_i (\boldsymbol{\varphi}^q_i \cdot \boldsymbol{\hat{n}}_i),
\end{equation}
such that:
\begin{equation}
    \frac{\mathrm{d}}{\mathrm{d} \tilde{t}} a^u_q + 
    \frac{1}{\pi k} \omega^2_q a^u_q = \pi_r \sum_i (\boldsymbol{\varphi}^q_i \cdot \boldsymbol{\hat{n}}_i).
\end{equation}

Let us consider the non-zero modes $\omega^2_q > 0$. If $\pi$ is such that $\omega^2_q/k \gg \pi$, then $a_q^u$ is severely damped and relaxes to zero on a timescale asymptotically fast in the rigid limit, $\pi \rightarrow 0$. We place ourselves at times larger than this typically transitory regime so that displacements along the non-zero modes can be safely neglected. On the other hand, the zero mode projections satisfy the following equation:
\begin{equation}
    \frac{\mathrm{d}}{\mathrm{d} \tilde{t}} a^u_q 
    = \pi_r \sum_i (\boldsymbol{\varphi}^q_i \cdot \boldsymbol{\hat{n}}_i).
\end{equation}

\section{Rigid theoretical framework}
 
Let $i$ and $j$ be two
agents in a pair. We may cast the geometrical inextensibility condition as:
\begin{equation} \label{eq:rigid_condition_1}
    \frac{\mathrm{d}}{\mathrm{d} t} \lVert \boldsymbol{x}_{i} - \boldsymbol{x}_{j} \rVert =
	\boldsymbol{\hat{e}}_{ij} \cdot 
	\left(\dot{\boldsymbol{x}}_{i} - \dot{\boldsymbol{x}}_{j} \right) = 0,
\end{equation}
where $\boldsymbol{\hat{e}_{ij}} = \frac{\boldsymbol{x}_{i} - \boldsymbol{x}_{j}}{\lVert \boldsymbol{x}_{i}- \boldsymbol{x}_{j} \rVert}$. If $j$ refers to a pinned site, then said equation simply reads:
\begin{equation}
    \boldsymbol{\hat{e}}_{ij} \cdot \dot{\boldsymbol{x}}_i = 0.
\end{equation}

There exist $M$ such equations, where $M$ is the number of edges of the system. We can put them in a compact form using bra-ket notation. Let $m$ denote the link between agents $i$ and $j$, then Eq. \ref{eq:rigid_condition_1} can be
written as:
\begin{equation}
    \braket{m | \mathcal{S} | \boldsymbol{\dot{x}}} = 0,
\end{equation}
where $\ket{\dot{\boldsymbol{x}}}$ is a particle-space ket satisfying:
\begin{equation}
    \braket{a | \dot{\boldsymbol{x}}} = \dot{\boldsymbol{x}}_a,
\end{equation}
and:
\begin{equation}
    \braket{m | \mathcal{S} | a} = \boldsymbol{\hat{e}}_{ij}^T (\delta_{ai} - \delta_{aj}),
\end{equation}
or, if $j$ refers to a pinned site:
\begin{equation}
    \braket{m | \mathcal{S} | a} = \boldsymbol{\hat{e}}_{ij}^T \delta_{ai},
\end{equation}
are the matrix elements of an $M \times 2N$ operator. Then, the set of all such equations can be expressed succinctly as:
\begin{equation} \label{eq:rigid_condition_braket}
    S \ket{\boldsymbol{\dot{x}}} = 0.
\end{equation}
The same matrix can be utilized to decompose the net force on each agent as the sum of the tensions along each edge. Indeed, we can always write:
\begin{equation}
    \boldsymbol{F}_i = \sum_{j \in V_i} f_{ij} \boldsymbol{\hat{e}}_{ij}.
\end{equation}
Let $m$ index the bonds $(j \rightarrow i)$. Then:
\begin{equation}
    \boldsymbol{F}_i = \sum_{m (j \rightarrow i)} \boldsymbol{\hat{e}}_m f_m.
\end{equation}
Now, let $m$ be an arbitrary bond $(p \rightarrow q)$ in the system. Each pair of particles is indexed once, with 
a particular bond orientation.
We must sum over all the (positive) contributions where $i$ is at the
end of this bond, and over all the (negative) contributions where $i$ is at the origin of the bond:
\begin{align}
    \boldsymbol{F}_i 
    &= \sum_{m} \boldsymbol{\hat{e}}_m \left(\delta_{iq} - \delta_{ip}\right) f_m, \\
    &= \sum_{m} \left(\braket{m | \mathcal{S} | i} \right)^T \braket{m | f}, \\
    &= \sum_{m} \braket{i | \mathcal{S}^T | m} \braket{m | f}, \\
    &= \braket{i | \mathcal{S}^T | f},
\end{align}
where we have used the completeness of the edge-space basis:
\begin{equation}
    \sum_m \ket{m} \bra{m} = \mathcal{I}.
\end{equation}
Finally, as $\boldsymbol{F}_{i} = \braket{i | \boldsymbol{F}}$:
\begin{equation} \label{eq:force_and_tensions}
    \ket{\boldsymbol{F}} = \mathcal{S}^T \ket{f}.
\end{equation}
Then, using \ref{eq:rigid_condition_braket} and \ref{eq:force_and_tensions} in \ref{ket_u_dot_before}, and noting that $\ket{\dot{\boldsymbol{u}}} = \ket{\dot{\boldsymbol{x}}}$:
\begin{equation} \label{eq:SSt_equation}
    \mathcal{S} \mathcal{S}^T \ket{f} = - \mathcal{S} \ket{\boldsymbol{\hat{n}}}.
\end{equation}
We can now decompose $\ket{f}$ in the eigenkets of $\mathcal{S} \mathcal{S}^T$:
\begin{equation}
    \ket{f} = \sum_q f_q \ket{\phi^q} = \sum_{q \in \mathfrak{Z}} f_q \ket{\phi^q} 
    + \sum_{q \notin \mathfrak{Z}} f_q \ket{\phi^q},
\end{equation}
where $\mathfrak{Z}$ is the set of zero modes of $\mathcal{S} \mathcal{S}^T$. We can then replace this expansion in \ref{eq:SSt_equation}:
\begin{equation}
    \sum_{q \notin \mathfrak{Z}} f_q \lambda_q^2 \ket{\phi^q} = - \mathcal{S} \ket{\boldsymbol{\hat{n}}},
\end{equation}
where $\lambda_q^2$ is the $q$-th eigenvalue of $\mathcal{S} \mathcal{S}^T$. Then:
\begin{equation}
    f_q = - \frac{\braket{ \phi^k | \mathcal{S} | \boldsymbol{\hat{n}}}}{\lambda_q^2} \ \forall q \notin \mathfrak{Z}.
\end{equation}
From this:
\begin{equation} \label{eq:ket_f_before}
    \ket{f} = \sum_{q \in \mathfrak{Z}} f_q \ket{\phi^q}
    - \sum_{q \notin \mathfrak{Z}} \frac{\braket{\phi^q | \mathcal{S} | \boldsymbol{\hat{n}}}}{\lambda_q^2} \ket{\phi^q}.
\end{equation}
At this point, it is important to realize that $\mathcal{S} \mathcal{S}^T$ shares a common set of eigenvalues with the non-dimensional dynamical matrix $\mathcal{S}^T \mathcal{S} = \frac{1}{k} \mathcal{M}$, with $k$ the spring constant. Let $\ket{\boldsymbol{\varphi}^p}$ be the $p$-th eigenket of the latter, then we 
can write:
\begin{equation}
    \mathcal{S}^T \mathcal{S} \ket{\boldsymbol{\varphi}^p} = \widetilde{\omega}^2_p \ket{\boldsymbol{\varphi}^p},
\end{equation}
where $\widetilde{\omega}^2_p = \frac{1}{k} \omega^2_p$. Then:
\begin{equation}
    \mathcal{S} \mathcal{S}^T \left(\mathcal{S} \ket{\boldsymbol{\varphi}^p} \right) 
    = 
    \widetilde{\omega}^2_p \left(\mathcal{S} \ket{\boldsymbol{\varphi}^p} \right).
\end{equation}
If $\ket{\boldsymbol{\varphi}^p} \notin \ker(\mathcal{S})$, then $\mathcal{S} \ket{\boldsymbol{\varphi}^p}$ is an eigenket of
$\mathcal{S} \mathcal{S}^T$ with eigenvalue $\widetilde{\omega}^2_p$. Then, every nonzero eigenvalue of $\mathcal{S}^T \mathcal{S}$ is also in $\mathcal{S} \mathcal{S}^T$, same with the zero eigenvalues associated with kets that are not in the kernel of $\mathcal{S}$, i.e $\mathcal{S} \ket{\boldsymbol{\varphi}^p} \in \ker{\mathcal{S}^T}$. We can also go the other way around, consider the
$q$-th eigenket of $\mathcal{S} \mathcal{S}^T$. Then:
\begin{equation}
    \mathcal{S} \mathcal{S}^T \ket{\phi^q} = \lambda^2_q \ket{\phi^q},
\end{equation}
such that:
\begin{equation}
    \mathcal{S}^T \mathcal{S} \left( \mathcal{S}^T \ket{\phi^q} \right)
    =
    \lambda^2_q \left( \mathcal{S}^T \ket{\phi^q} \right).
\end{equation}
If $\ket{\phi^q} \notin \ker(\mathcal{S}^T)$, then $\mathcal{S}^T \ket{\phi^q}$ is an eigenket of
$\mathcal{S}^T \mathcal{S}$ with eigenvalue $\lambda^2_q$. Then, every nonzero eigenvalue of 
$\mathcal{S} \mathcal{S}^T$ is also in $\mathcal{S}^T \mathcal{S}$, same with the zero eigenvalues associated with kets that are not in the kernel of $\mathcal{S}^T$, i.e $\mathcal{S}^T \ket{{\phi}^q} \in \ker{\mathcal{S}}$.

With this, we conclude that the two matrices have exactly the same nonzero eigenvalues, and noting that the two conditions for shared zero eigenvalues conflict, they do not share any zero eigenvalues. We can then write the following:
\begin{align}
    \mathcal{S}^T \ket{\phi^q} &= \alpha_q \ket{\boldsymbol{\varphi}^q}, \\
    \mathcal{S} \ket{\boldsymbol{\varphi}^q} &= \beta_q \ket{\phi^q},
\end{align}
so that:
\begin{align}
    \mathcal{S} \mathcal{S}^T \ket{\phi^q} &= \alpha_q \beta_q \ket{\phi^q}, \\
    \mathcal{S}^T \mathcal{S} \ket{\boldsymbol{\varphi}^q} &= \beta_q \alpha_q \ket{\boldsymbol{\varphi}^q}.
\end{align}
We have established that $\widetilde{\omega}^2_q = \lambda^2_q$ for eigenkets not in the kernel of $\mathcal{S}$ or $\mathcal{S}^T$. With this, we conclude that:
\begin{align}
    \mathcal{S}^T \ket{\phi^q} &= \lambda_q \ket{\boldsymbol{\varphi}^q}, \label{eq:condition_relating_ST} \\
    \mathcal{S} \ket{\boldsymbol{\varphi}^q} &= \widetilde{\omega}_q \ket{\phi^q},
\end{align}
which works even if $\mathcal{S}^T \ket{\phi^q} = 0$ or $\mathcal{S} \ket{\boldsymbol{\varphi}^q} = 0$ as in that case clearly $\lambda_q = 0$ and $\widetilde{\omega}_q = 0$ respectively.

Finally, using \ref{eq:condition_relating_ST} in \ref{eq:ket_f_before}:
\begin{equation}
    \ket{\boldsymbol{F}} = \mathcal{S}^T \ket{f}
    = - \sum_{q \notin \mathfrak{F}} \braket{\boldsymbol{\varphi}^q | \boldsymbol{\hat{n}}} \ket{\boldsymbol{\varphi}^q},
\end{equation}
where $\mathfrak{F}$ is the set of zero modes of $S^T S$, and using the resolution of the identity:
\begin{equation} \label{eq:explicit_force}
    \ket{\boldsymbol{F}} = - \ket{\boldsymbol{\hat{n}}} + \sum_{q \in \mathfrak{F}} \braket{\boldsymbol{\varphi}^q | \boldsymbol{\hat{n}}} \ket{\boldsymbol{\varphi}^q}.
\end{equation}

We will derive an equation that will be useful to describe the dynamics of the system:
\begin{align}
    \braket{\boldsymbol{\varphi}^q | \dot{\boldsymbol{x}}}
    &=\sum_{p \in \mathfrak{F}} \braket{\boldsymbol{\varphi}^p | \boldsymbol{\hat{n}}} \braket{\boldsymbol{\varphi}^q | \boldsymbol{\varphi}^p}, \\
    &= \braket{\boldsymbol{\varphi}^q | \boldsymbol{\hat{n}}}, \label{eq:useful1}
\end{align}
where we have used the orthonormality of the eigenvector basis.

\section{Fokker-Planck equation}

Consider a system described by a set of $W$ random variables ${y_i}$ evolving under the following stochastic differential equation:
\begin{equation}
    \dot{y}_i(t) = L_i(y_1,\dots,y_W | t) + \sigma_i \xi_i(t),
\end{equation}
where $\xi_i$ is a delta-correlated Gaussian white noise such that:
\begin{align}
    \langle \xi_i(t) \rangle &= 0, \\
    \langle \xi_i(t) \xi_j(t+\tau) \rangle &= \delta_{ij} \delta(\tau).
\end{align}

Let $Q(y_1,\dots,y_W  | t)$ be the probability density function of a particular configuration of the system at time $t$. The Fokker-Planck equation describing the spatiotemporal evolution of $Q$ reads:
\begin{equation}
    \frac{\partial Q}{\partial t }
    = - \sum_i \frac{\partial}{\partial y_i} \left(L_i Q \right) + \sum_i \frac{\sigma_i^2}{2} \frac{\partial^2 Q}{\partial y_i^2}.
\end{equation}
In our particular case, $W=3N$. Furthermore, $\sigma_i=0$ for all the equations describing the particle positions. From the dynamical equations, the Fokker-Planck equation reads:
\begin{equation}
    \frac{\partial}{\partial t} Q(\boldsymbol{x}_1,\dots,\boldsymbol{x}_N | \theta_1, \dots, \theta_N | t)
    = - \sum_{i=1}^N \boldsymbol{\nabla}_{\boldsymbol{x}_i} \cdot \left((\boldsymbol{\hat{n}}_i + \boldsymbol{F}_i) Q\right)
    - \frac{1}{\pi_r} \sum_{i=1}^N \frac{\partial}{\partial \theta_i} \left((\boldsymbol{\hat{n}}^\perp_i \cdot \boldsymbol{F}_i) Q \right)
    + \frac{\pi_{\theta}}{\pi_r} \sum_{i=1}^N  \frac{\partial^2 Q}{\partial \theta_i^2}.
\end{equation}
And from \ref{eq:explicit_force}:
\begin{equation}
    \frac{\partial Q}{\partial t} =  - \sum_{i,q} \boldsymbol{\nabla}_{\boldsymbol{x}_i} \cdot
    \left(  \langle\boldsymbol{\varphi}^q | \boldsymbol{\hat{n}}\rangle  
    \boldsymbol{\varphi}_i^q Q\right) 
    +  \sum_i \frac{\partial}{\partial \theta_i}
    \left(\frac{1}{\pi_r} \frac{\partial V}{\partial \theta_i} Q \right) 
    + \frac{\pi_{\theta}}{\pi_r} \sum_i \frac{\partial^2 Q}{\partial \theta_i^2}.
    \label{eq:fp_full}
\end{equation}
Now let us assume that the modes do not depend on the particle positions. This means that we can integrate Eq. \ref{eq:fp_full} with respect to them:
\begin{equation}
    \frac{\partial \mathcal{Q}}{\partial t} =  
    \sum_i \frac{\partial}{\partial \theta_i}
    \left(\frac{1}{\pi_r} \frac{\partial V}{\partial \theta_i} \mathcal{Q} \right) 
    + \frac{\pi_{\theta}}{\pi_r} \sum_i \frac{\partial^2 \mathcal{Q}}{\partial \theta_i^2},
    \label{eq:fp_reduced}
\end{equation}
where $\mathcal{Q} = \int Q \, \mathrm{d}^N \boldsymbol{x}$. To find a steady-state solution to this equation, we may write:
\begin{equation}
    0 = \sum_i \frac{\partial}{\partial \theta_i} \left[\left(\frac{1}{\pi_r} \frac{\partial V}{\partial \theta_i} \mathcal{Q} \right) + \frac{\pi_{\theta}}{\pi_r} \frac{\partial \mathcal{Q}}{\partial \theta_i} \right],
\end{equation}
which has a solution given by the Gibbs measure:
\begin{equation}
    \mathcal{Q} = \frac{1}{\mathcal{Z}} e^{- \beta V}, \label{eq:gibbs}
\end{equation}
where $\beta = 1/\pi_{\theta}$ is the inverse temperature and $\mathcal{Z}$ is the partition function:
\begin{equation}
    \mathcal{Z} = \int e^{- \beta V} \, \mathrm{d}^N \theta.
\end{equation}
If some modes do depend on the particle positions, we can state the following. If $\pi_r \ll 1$, it means that particles quickly reorient their polarity vectors to wherever they're moving. When $\pi_r \rightarrow 0$, the timescale of this reorientation is instantaneous in comparison to the timescale of particle displacements, i.e. mode evolution. Then, we may assume an adiabatic evolution where for each spatial configuration of the system, the particles can reach a steady-state polarity angle distribution before changing the spatial configuration. In essence, the $\theta$ variables evolve while the positional degrees of freedom are frozen, i.e. the slow structure parameters enslave the polarity vectors. Then, we can still integrate Eq. \ref{eq:fp_full} with respect to the spatial configurations that leave the modes unchanged,
and consider the other modes fixed such that we don't have spatial derivatives anymore. In the end, we get the same Eq. \ref{eq:fp_reduced} and solution as in Eq. \ref{eq:gibbs}, with the potential and the partition function depending on the structure parameters of the system, which then evolve according to the steady-state configuration of the system. This procedure, and the $\pi_r$ expansion, can be done formally (See section \ref{section:rotational_system} for an example).

\section{Statistics of the adiabatic approximation}

\subsection*{General formulation}

We can calculate the partition function as:
\begin{equation} \label{eq:partition_function_1}
    \mathcal{Z} = \int_0^{2 \pi} \exp \left(\beta \left( 
    \frac{1}{2} \sum_q \left(\braket{\boldsymbol{\varphi}^q | \boldsymbol{\hat{n}}} \right)^2
    + \sqrt{N} \sum_q h_q  \braket{\boldsymbol{\varphi}^q | \boldsymbol{\hat{n}}} \right) \right)
    \, \mathrm{d}^N \theta.
\end{equation}
We note that, for a globally extended mode, $\braket{\boldsymbol{\varphi}^q | \boldsymbol{\hat{n}}}$ 
is $\mathcal{O}(\sqrt{N})$. We will write:
\begin{equation}
    \exp \left( \frac{\beta}{2N} \left( \sqrt{N} \braket{\boldsymbol{\varphi}^q | \boldsymbol{\hat{n}}} \right)^2 \right) 
    =
    \sqrt{\frac{N \beta}{2 \pi}} \int_{-\infty}^{\infty}
    \exp \left( - \frac{N\beta}{2} \mu_q^2 + \left(\beta \sqrt{N} \braket{\boldsymbol{\varphi}^q| \boldsymbol{\hat{n}}} \right) \mu_q
    \right) \, \mathrm{d} \mu_q,
\end{equation}
so that:
\begin{equation}
    \mathcal{Z} = \left( \frac{N \beta}{2 \pi} \right)^{\frac{M}{2}} 
    \int_{- \infty}^{\infty}
    \exp \left(\frac{N\beta}{2} \sum_q \mu_q^2 \right) I(\mu_1, \dots, \mu_M) \, \mathrm{d}^{M} \mu,
\end{equation}
where:
\begin{equation}
I = \int_{0}^{2\pi} \exp \left(\beta \sum_k \left( \sqrt{N} \braket{\boldsymbol{\varphi}^q|\boldsymbol{\hat{n}}} \right) 
\left(\mu_q + h_q \right) \right) \, \mathrm{d}^N \theta.
\end{equation}
We note that for any of the global modes we can write:
\begin{equation}
    \ket{\boldsymbol{\varphi}^q} = \frac{1}{\sqrt{N}} \sum_i r_{qi} 
    \left( \cos(\phi_{qi}) \boldsymbol{\hat{x}} + \sin(\phi_{qi}) \boldsymbol{\hat{y}} \right),
\end{equation}
then:
\begin{equation}
    I 
    = \int_0^{2 \pi} \prod_i \left[\exp \left(a_i \cos(\theta_i) + b_i \sin(\theta_i) \right) \, \mathrm{d} \theta_i \right],
\end{equation}
where:
\begin{align}
    a_i &= \sum_q (\mu_q + h_q) r_{qi} \cos(\phi_{qi}), \\
    b_i &= \sum_q (\mu_q + h_q) r_{qi} \sin(\phi_{qi}).
\end{align}
Thus:
\begin{equation}
    I = (2 \pi)^N \prod_i I_0 \left(\beta \sqrt{a_i^2 + b_i^2} \right).
\end{equation}
This expression can be rewritten in a more tractable way. Note that:
\begin{align}
    a_i^2 + b_i^2 &= \sum_{q,l} (\mu_q + h_q)(\mu_l + h_l) r_{qi} r_{li} 
    \left(\cos(\phi_{qi})\cos(\phi_{li}) + \sin(\phi_{qi})\sin(\phi_{li}) \right), \\
    &= \sum_{q,l} (\mu_q + h_q)(\mu_l + h_l) r_{qi} r_{li} \cos(\phi_{qi} - \phi_{li}), \\
    &= N\sum_{q,l} (\mu_q + h_q)(\mu_l + h_l) \left(\boldsymbol{\varphi}^q_i \cdot \boldsymbol{\varphi}^l_i \right),
\end{align}
then we can define:
\begin{equation}
    \mathcal{D}_i = \left(N\sum_{q,l} (\mu_q + h_q)(\mu_l + h_l) \left(\boldsymbol{\varphi}^q_i \cdot \boldsymbol{\varphi}^l_i \right) \right)^{1/2},
\end{equation}
so that, up to a multiplicative constant:
\begin{equation}
    \mathcal{Z} = \int_{- \infty}^{\infty}
    \exp \left(- N \beta f (\mu_1, \dots, \mu_M) \right) \, \mathrm{d}^{M} \mu,
\end{equation}
where:
\begin{equation}
    f = \frac{1}{2} \sum_q \mu_q^2 - \frac{1}{N \beta} \sum_i \ln \left(I_0(\beta \mathcal{D}_i(\mu_1, \dots, \mu_M)) \right).
\end{equation}

\subsection*{Thermodynamic limit: Saddle-point approximation}

$\mathcal{D}_i$ is $\mathcal{O}(1)$, hence the free energy per particle in the thermodynamic limit is given by the minimum of $f$. Said minimum satisfies, for every mode k:
\begin{equation}
    \widetilde{\mu}_q = 
    \left. \frac{1}{N \beta} \sum_i \frac{\partial \ln \left(I_0(\beta \mathcal{D}_i)\right)}{\partial \mu_q} \right\vert_{\substack{h_l=0 \ \forall l \\ \mu_l = \widetilde{\mu}_l \ \forall l}}.
\end{equation}
We are interested in calculating the mean values of the projections of each mode over the polarity field. From the partition function definition in \ref{eq:partition_function_1}:
\begin{equation}
    \braket{\braket{\boldsymbol{\varphi}^q | \boldsymbol{\hat{n}}}} = \left. \frac{1}{\sqrt{N} \beta}
    \frac{\partial \ln(\mathcal{Z})}{\partial h_q} \right\vert_{h_l=0 \ \forall l}.
\end{equation}
Note that $\mathcal{Z}$ depends on $h_q$ only through $f$. Then:
\begin{equation}
    \langle \braket{\boldsymbol{\varphi}^q | \boldsymbol{\hat{n}}} \rangle
    =
    \left. \frac{1}{\mathcal{Z}} \frac{1}{\sqrt{N} \beta}
    \int_{-\infty}^{\infty} \exp \left(-N \beta f \right) 
    \left( - N \beta \frac{\partial f}{\partial h_q} \right) \, \mathrm{d}^{M} \mu 
    \right\vert_{h_l = 0 \ \forall l}.
\end{equation}
Thus:
\begin{equation}
    \frac{\langle \braket{\boldsymbol{\varphi}^q | \boldsymbol{\hat{n}}} \rangle}{\sqrt{N}}
    = 
    \left. \frac{1}{\mathcal{Z}}
    \int_{-\infty}^{\infty} \exp \left(-N \beta f \right) \left(\sum_i 
    \frac{\partial \ln \left(I_0(\beta \mathcal{D}_i)\right)}{\partial h_q} \right) \, \mathrm{d}^M \mu 
    \right\vert_{h_l = 0 \ \forall l}.
\end{equation}
From the explicit form of $\mathcal{D}_i$:
\begin{equation}
    \frac{\partial \ln\left(I_0( \beta\mathcal{D}_i) \right)}{\partial h_q} = \frac{\partial \ln \left(I_0(\beta \mathcal{D}_i) \right)}{\partial \mu_q}.
\end{equation}
Finally, ignoring fluctuations up to first order in the saddle point approximation:
\begin{equation}
    \frac{\langle \braket{\boldsymbol{\varphi}^q | \boldsymbol{\hat{n}}} \rangle}{\sqrt{N}}
    = \left. \frac{1}{N \beta} \sum_i \frac{\partial \ln \left(I_0(\beta \mathcal{D}_i)\right)}{\partial \mu_q} \right\vert_{\substack{h_l=0 \ \forall l \\ \mu_l = \widetilde{\mu}_l \ \forall l}}.
\end{equation}
Then, the values:
\begin{equation}
    \widetilde{\mu}_q 
    = \frac{\langle \braket{\boldsymbol{\varphi}^q | \boldsymbol{\hat{n}}} \rangle}{\sqrt{N}},
\end{equation}
define a minimum of the free energy.

\subsection*{Phase transition}

Let us consider the limit where the order parameters are very small, then we can use the approximation:
\begin{equation} \label{eq:ln_approx}
    \ln(I_0(x)) \approx \frac{x^2}{4},
\end{equation}
such that:
\begin{align}
    f
    &=
    \frac{1}{2} \sum_q \mu_q^2
    - \frac{\beta}{4} \sum_{i,q,l} \mu_q \mu_l \left(\boldsymbol{\varphi}^q_i \cdot \boldsymbol{\varphi}^l_i \right), \\
    &=
    \frac{1}{2} \sum_q \mu_q^2
    - \frac{\beta}{4} \sum_{q,l} \mu_q \mu_l \braket{\boldsymbol{\varphi}^q | \boldsymbol{\varphi}^l}.
\end{align}
Now, as the modes are orthonormal:
\begin{equation}
    f
    =
    \frac{1}{2} \left(1 - \frac{\beta}{2} \right) \sum_q \mu_q^2.
\end{equation}
Then, there is a continuous transition at $\beta=2$ for every mode, with $\beta > 2$ corresponding to the ordered phase.

\section{Translational system}

For a PBC system, the zero modes of the system correspond to global translations, which can be represented with an orthogonal basis as follows:
\begin{align}
    \boldsymbol{\varphi}^x_i &= \frac{1}{\sqrt{N}} \boldsymbol{\hat{x}}, \\
    \boldsymbol{\varphi}^y_i &= \frac{1}{\sqrt{N}} \boldsymbol{\hat{y}}.
\end{align}
From this:
\begin{align}
    V &= - \frac{1}{2} \left(\frac{1}{N} \left(\sum_i \boldsymbol{\hat{n}}_i \cdot \boldsymbol{\hat{x}} \right)^2 
    + \frac{1}{N}\left( \sum_i \boldsymbol{\hat{n}}_i \cdot \boldsymbol{\hat{y}} \right)^2 \right), \\
    &= - \frac{1}{2N} \left(\sum_i \boldsymbol{\hat{n}}_i \right)^2, \\
    &= - \frac{1}{2N} \sum_{ij} \left(\boldsymbol{\hat{n}}_i \cdot \boldsymbol{\hat{n}_j} \right).
\end{align}
As the modes do not depend on the position of the structure, the conditions described in the main text are fulfilled, and the reduced probability distribution $\mathcal{Q}$ of the state of the system follows a Gibbs weight. The Landau free energy reads:
\begin{equation}
    f = \frac{1}{2} \left(\boldsymbol{\mu}_P^2 \right) - \frac{1}{N \beta} \log (I_0(\beta \lVert \boldsymbol{\mu}_P \rVert)),
\end{equation}
where:
\begin{equation}
    \boldsymbol{\mu}_P = \mu_x \boldsymbol{\hat{x}} + \mu_y \boldsymbol{\hat{y}} = \frac{1}{N} \left \langle \sum_i \boldsymbol{\hat{n}}_i \right \rangle.
\end{equation}

\subsection*{Order parameter critical exponent}

In this case, we can find the order parameter critical exponent by continuing the expansion from Eq. \ref{eq:ln_approx} up to the next order:
\begin{equation}
    \log (I_0(x)) \approx \frac{x^2}{4} - \frac{x^4}{64}.
\end{equation}
Then:
\begin{equation}
    f = \frac{1}{2} \left(1 - \frac{\beta}{2} \right) \mu_P^2 + \frac{\beta^3}{64} \mu_P^4,
\end{equation}
such that in the ordered phase ($\beta$ > 2), the following holds:
\begin{equation}
    \mu_P^2 = \frac{16}{\beta^3} \left(\frac{\beta}{2} - 1 \right),
\end{equation}
then:
\begin{equation}
    \mu_P^2 = 16 \pi_{\theta}^2 \left(\frac{1}{2} - \pi_{\theta} \right),
\end{equation}
and finally:
\begin{equation}
    \mu_P \sim (\pi_{\theta}^c - \pi_{\theta})^{1/2}.
\end{equation}

\section{Rotational system} \label{section:rotational_system}

As the particles rotate, they describe circumferences centered at the pinning point. Then, we can write the rotational
mode as:
\begin{equation}
    \boldsymbol{\varphi}_i^{\phi} = \frac{1}{\sqrt{I}} \boldsymbol{r}^{\perp}_i,
\end{equation}
where $\boldsymbol{r}_i$ is the particle's position with respect to the pinning point, and  
$I = \sum_i r_i^2$ is the rotational moment of inertia. 

We can express the velocity of each particle as:
\begin{equation}
    \dot{\boldsymbol{x}}_i = \dot{\phi} \boldsymbol{r}_{i}^{\perp},
\end{equation}
where $\phi$ is the angle describing the rotation of the structure. The evolution of $\phi$ can be described by using
Eq. \ref{eq:useful1}:
\begin{align}
    \braket{\boldsymbol{\varphi}^{\phi} | \dot{\boldsymbol{x}}} &=
    \braket{\boldsymbol{\varphi}^{\phi} | \boldsymbol{\hat{n}}}, \\
    \dot{\phi} \frac{1}{\sqrt{I}} \sum_i \boldsymbol{r}_{i}^{\perp} \cdot \boldsymbol{r}_{i}^{\perp}
    &= \frac{1}{\sqrt{I}} \sum_i \boldsymbol{r}_{i}^{\perp} \cdot \boldsymbol{\hat{n}}_i,
\end{align}
then:
\begin{equation}
    \dot{\phi} = \frac{1}{I} \sum_i \boldsymbol{r}_{i}^{\perp} \cdot \boldsymbol{\hat{n}}_i.
\end{equation}
In the laboratory reference frame, as the particles rotate, the rotational mode evolves in time because the position of each particle is changing. But if we choose a reference frame that rotates with the structure, the mode remains invariant in time. This allows us to find an approximate solution to the Fokker-Planck equation in a controlled way. The state of the system can be characterized by $\phi$, the global rotational angle, and the set of  $\theta_i' = \theta_i - \phi$. We can write the dynamics of the system as follows:
\begin{align}
    \dot{\phi} &= \frac{1}{I} \sum_i \boldsymbol{r}_{i}^{\perp} \cdot \boldsymbol{\hat{n}}_i, \\
    \dot{\theta_i'} &= - \frac{1}{\pi_r} \frac{\partial V}{\partial \theta_i'} - \dot{\phi} + \sqrt{\frac{2 \pi_{\theta}}{\pi_r}} \xi_i,
\end{align}
where:
\begin{equation}
    V = - \frac{1}{2I} \left(\sum_i \boldsymbol{r}^{\perp}_{i} \cdot \boldsymbol{\hat{n}}_i \right)^2.
\end{equation}
We can first study the behavior of the system in the noiseless, $\pi_\theta = 0$, case. We may note that:
\begin{align}
    \frac{\partial V}{\partial \theta_i'} &= -\frac{1}{I} \sum_j \boldsymbol{r}_{j}^{\perp} \cdot \boldsymbol{\hat{n}_j}
    \left(\boldsymbol{r}_{i}^{\perp} \cdot \boldsymbol{\hat{n}}_i^{\perp} \right), \\
    &= -\dot{\phi} \left(\boldsymbol{r}_{i} \cdot \boldsymbol{\hat{n}}_i \right).
\end{align}
Then the steady-state, zero-noise solution satisfies:
\begin{equation}
    \dot{\theta_i'} =  \left(\frac{1}{\pi_r} \left(\boldsymbol{r}_{i} \cdot \boldsymbol{\hat{n}}_i \right) - 1\right) \dot{\phi} = 0,
\end{equation}
and hence:
\begin{equation}
    \left(\boldsymbol{r}_{i} \cdot \boldsymbol{\hat{n}}_i \right) = \pi_r,
\end{equation}
such that, as $\pi_r \rightarrow 0$, the polarity vectors become perpendicular to the positional vectors from the
pinning point. As a consequence, if $\pi_r > r_{\text{min}}$, then this rotational solution does not exist. 

To find the explicit form of $\dot{\phi}$, we utilize the expansion:
\begin{equation}
    r_i^2 \hat{\boldsymbol{n}_i} = \left(\boldsymbol{r}_{i} \cdot \boldsymbol{\hat{n}}_i \right) \boldsymbol{r}_{i}
    + \left(\boldsymbol{r}_{i}^{\perp} \cdot \boldsymbol{\hat{n}}_i \right) \boldsymbol{r}_{i}^{\perp},
\end{equation}
such that, squaring both sides:
\begin{equation}
    r_i^2 = \left(\boldsymbol{r}_{i} \cdot \boldsymbol{\hat{n}}_i \right)^2 + \left(\boldsymbol{r}_{i}^{\perp} \cdot \boldsymbol{\hat{n}}_i \right)^2,
\end{equation}
and hence:
\begin{align}
    \boldsymbol{r}_{i}^{\perp} \cdot \boldsymbol{\hat{n}}_i &= \pm \sqrt{r_i^2 - \left(\boldsymbol{r}_{i} \cdot \boldsymbol{\hat{n}}_i \right)^2 }, \\
    &= \sqrt{r_i^2 - \pi_r^2},
\end{align}
as such, considering the solution where every particle rotates in the same direction:
\begin{equation}
    \dot{\phi} = \frac{1}{I} \sum_i \sqrt{r_i^2 - \pi_r^2}.
\end{equation}
Coming back to the $\pi_{\theta} \neq 0$, we study the Fokker-Planck equation:
\begin{equation}
    \frac{\partial Q}{\partial t} = - \frac{\partial }{\partial \phi} \left[\dot{\phi} Q\right]
    + \sum_i \frac{\partial}{\partial \theta_i'} \left[\left(\frac{1}{\pi_r} \frac{\partial V}{\partial \theta_i'} + \dot{\phi} \right) Q \right] + \frac{\pi_{\theta}}{\pi_r} \sum_i \frac{\partial^2 Q}{\partial \theta_i'^2}.
\end{equation}
Integrating with respect to $\phi$ for all orientations:
\begin{equation}
    \frac{\partial \mathcal{Q}}{\partial t} =
    \sum_i \frac{\partial}{\partial \theta_i'} \left[\left(\frac{1}{\pi_r} \frac{\partial V}{\partial \theta_i'} + \dot{\phi} \right) \mathcal{Q} \right] + \frac{\pi_{\theta}}{\pi_r} \sum_i \frac{\partial^2 \mathcal{Q}}{\partial \theta_i'^2}.
    \label{eq:fp_rotational}
\end{equation}
Now
\begin{align}
    \frac{\partial V}{\partial \theta_i'} &\sim R \dot{\phi}, \\
    \dot{\phi} &\sim \frac{1}{R},
\end{align}
where $R$ is the characteristic agent distance from the center of mass. If $\pi_r \ll R$, we may ignore the $\dot{\phi}$ term. More precisely, we can perform a series expansion of $\mathcal{Q}$ as follows:
\begin{equation}
    \mathcal{Q} = \mathcal{Q}^{(0)} + \pi_r \mathcal{Q}^{(1)} + \dots
\end{equation}
Using this expansion in Eq. \ref{eq:fp_rotational}, at order $\mathcal{O}(\pi_r^{-1})$, the $\dot{\phi}$ term is not present, and we recover the steady-state solution given by the Gibbs factor. The same solution can be obtained via the adiabatic assumption described in the main text. In other words, such a procedure already considers $\pi_r \rightarrow 0$ from the very beginning. For any higher order $\mathcal{O}(\pi_r^p)$ the steady-state solution satisfies the following equation:
\begin{equation} \label{eq:fp_rt_2}
    \boldsymbol{\nabla} \cdot \left(e^{-\beta V} \boldsymbol{\nabla} \left(e^{\beta V} \mathcal{Q}^{(p)} \right) \right) = \beta
    \sum_i \frac{\partial}{\partial \theta'_i} \left( \dot{\phi} \mathcal{Q}^{(p-1)} \right).
\end{equation}
Then, the Landau free energy characterizing the rotational system is:
\begin{equation}
    f = \frac{1}{2} \mu_{\phi}^2
    - \frac{1}{N \beta} \sum_{i} \log \left(I_0 \left(\beta \sqrt{N/I} r_i \lvert \mu_{\phi} \rvert\right)\right),
\end{equation}
where:
\begin{equation}
    \mu_{\phi} = \sqrt{\frac{I}{N}} \langle \dot{\phi} \rangle.
\end{equation}

Following the procedure that we have described here, there is no need to evolve the state of the system as our angles are measured with respect to $\phi$. Had we decided to implement the adiabatic assumption in the lab frame, we would need to evolve $\phi$ as follows:
\begin{equation}
    \frac{\mathrm{d} \phi}{\mathrm{d} t} = \langle \dot{\phi} \rangle,
\end{equation}
and rotate the positions $\boldsymbol{r}_{i}$ accordingly, then iterate the process again.

\section{Auxetic squares system}

We may first note that, as the auxetic angle evolves, all the particles describe circular trajectories. This can be understood by considering the first layer of auxetic squares, where the evolution of the auxetic angle leads the four free particles in each square to describe a circle around the pinned corner. By the symmetry of the system, all the particles move in circles with different radii and centers. Then, for any particle $i$, we can write the auxetic mode as:
\begin{equation}
    \boldsymbol{\varphi}_i^{\gamma} = \frac{1}{\sqrt{\tilde{I}}} \tilde{\boldsymbol{r}}^{\perp}_i,
\end{equation}
where $\tilde{\boldsymbol{r}}_{i}$ is the particle's position with respect to its rotational center, and  
$\tilde{I} = \sum_i \tilde{r}_i^2$ is the auxetic moment of inertia. 

We can express the velocity of each particle as:
\begin{equation}
    \dot{\boldsymbol{x}}_i = \dot{\gamma} \boldsymbol{\tilde{r}}_{i}^{\perp},
\end{equation}
where $\gamma$ is the angle describing the auxetic deformation of the structure. The evolution of $\gamma$ can be described by using Eq. \ref{eq:useful1}:
\begin{align}
    \braket{\boldsymbol{\varphi}^{\gamma} | \dot{\boldsymbol{x}}} &=
    \braket{\boldsymbol{\varphi}^{\gamma} | \boldsymbol{\hat{n}}}, \\
    \dot{\phi} \frac{1}{\sqrt{\widetilde{I}}} \sum_i \boldsymbol{\tilde{r}}_{i}^{\perp} \cdot \boldsymbol{\tilde{r}}_{i}^{\perp}
    &= \frac{1}{\sqrt{\widetilde{I}}} \sum_i \boldsymbol{\tilde{r}}_{i}^{\perp} \cdot \boldsymbol{\hat{n}}_i,
\end{align}
then:
\begin{equation}
    \dot{\gamma} = \frac{1}{\widetilde{I}} \sum_i \boldsymbol{\tilde{r}}_{i}^{\perp} \cdot \boldsymbol{\hat{n}}_i.
\end{equation}
We can repeat the same process as in the rotational structure, now choosing to measure angles with respect to $\gamma$, as in such a rotational frame the mode remains invariant in time, with: 
\begin{equation}
    V = - \frac{1}{2\widetilde{I}} \left(\sum_i \boldsymbol{\tilde{r}}^{\perp}_i \cdot \boldsymbol{\hat{n}}_i \right)^2.
\end{equation}
In the zero-noise case, the steady-state solution satisfies:
\begin{equation}
    (\boldsymbol{\tilde{r}}_{i} \cdot \boldsymbol{\hat{n}}_i) = \pi_r,
\end{equation}
which constrains the maximum $\pi_r$ value such that purely auxetic motion is not possible. Given the regular spacing
between particles, the smaller rotation radii will correspond to the centermost particles, with $\tilde{r}_{\text{min}} = \sqrt{2}/2$. Similarly, we can find the zero-noise dynamics of the auxetic angle analogously to the rotational case:
\begin{equation}
    \dot{\gamma} = \frac{1}{\widetilde{I}} \sum_i \sqrt{\tilde{r}_i^2 - \pi_r^2}.
\end{equation}

Treating the $\pi_{\theta}>0$ case, we arrive at the same conclusion as in the purely rotational case:
we can write $\mathcal{Q}$ for this system as a Gibbs measure in the $\pi_r \ll 1$ limit.

In the thermodynamic limit, the mean temporal evolution of the auxetic angle $\gamma$ is given by the minimum
of the free energy:
\begin{equation}
    f = \frac{1}{2} \mu_{\gamma}^2
    - \frac{1}{N \beta} \sum_{i} \log \left[ I_0 \left(\beta \sqrt{N/\widetilde{I}} \tilde{r}_i \lvert \mu_{\gamma} \rvert\right)\right],
\end{equation}
where:
\begin{equation}
    \mu_{\gamma} = \sqrt{\frac{\widetilde{I}}{N}} \langle \dot{\gamma} \rangle.
\end{equation}

Following the procedure that we have described here, there is no need to evolve the state of the system as our angles are measured with respect to $\gamma$. Had we decided to implement the adiabatic assumption in the lab frame, we would need to evolve $\gamma$ as follows:
\begin{equation}
    \frac{\mathrm{d} \gamma}{\mathrm{d} t} = \langle \dot{\gamma} \rangle,
\end{equation}
and rotate the positions $\boldsymbol{\tilde{r}}_{i}$ accordingly, then iterate the process again.

It is possible to write analytically the particle circumference radii and find $\widetilde{I}$. We may separate particles in square vertices and square centers, such that:
\begin{equation}
    \tilde{I} = \sum_{k=1}^L \Xi^v(k) + \sum_{k=1}^L \Xi^c(k),
\end{equation}
where L is the number of layers, and the decomposition reads:
\begin{equation}
    \Xi^c(k) =
    \begin{cases}
    4 \left(k \sqrt{2}/2 \right)^2 + 4 \left(k/2 \right)^2 + 8 \sum_{q=1}^{k/2-1} \left(\sqrt{k^2 + (2q)^2}/2 \right)^2 & k \text{ even}, \\
    4 \left(k \sqrt{2}/2 \right)^2 + 8 \sum_{q=1}^{(k-1)/2} \left(\sqrt{k^2 + (2q-1)^2}/2 \right)^2 & k \text{ odd}, 
    \end{cases}
\end{equation}
and:
\begin{equation}
    \Xi^v(k) =
    \begin{cases}
    8 \left(k/2 \right)^2 + 16 \sum_{q=1}^{k/2-1} \left(\sqrt{(k/2)^2 + q^2} \right)^2 + 12 \left(\sqrt{2(k/2)^2} \right)^2 & k \text{ even}, \\
    8 \left(k - (k-1)/2 \right)^2 + 16 \sum_{q=1}^{k-(k-1)/2-1} \left(\sqrt{(k-(k-1)/2)^2 + q^2} \right)^2
    + 4 \left(\sqrt{2(k-(k-1)/2)^2} \right)^2 & k \text{ odd}.
    \end{cases}
\end{equation}

\section{Translational-Rotational system}

This system is characterized by three modes:
\begin{align}
    \boldsymbol{\varphi}_i^{x} &= \frac{1}{\sqrt{N}} \boldsymbol{\hat{x}}, \\
    \boldsymbol{\varphi}_i^{y} &= \frac{1}{\sqrt{N}} \boldsymbol{\hat{y}}, \\
    \boldsymbol{\varphi}_i^{\phi} &= \frac{1}{\sqrt{I}} \boldsymbol{r}^{\perp}_i.
\end{align}
It is trivial to show that they form an orthonormal basis, that is $\braket{\boldsymbol{\varphi}^k | \boldsymbol{\varphi}^q} = \delta_{kq}$. In this case, the relevant variables are the position of the center of mass, the rotation angle of the structure, and the polarization of each agent. We must write the dynamical equation for each one of them. From the definition of the center of mass, we can find the equation governing its 
temporal evolution as follows:
\begin{align}
    \dot{\boldsymbol{X}} &= \frac{1}{N} \sum_i \dot{\boldsymbol{x}}_i, \\
    &= \frac{1}{N} \sum_i \left(\boldsymbol{\hat{n}}_i + \boldsymbol{F}_{i} \right), \\
    &= \frac{1}{N} \sum_i \boldsymbol{\hat{n}}_i,
\end{align}
where the sum of all the inter-particle forces vanishes. We will now find the rotational speed. We can express the velocity of each particle as:
\begin{equation}
    \dot{\boldsymbol{x}}_i = \dot{\boldsymbol{X}} + \dot{\phi} \boldsymbol{r}_{i}^{\perp},
\end{equation}
where $\boldsymbol{r}_{i}$ is the position of particle $i$ with respect to the center of mass, and $\phi$ is the angle describing the rotation of the structure. The evolution of $\phi$ can be described by using Eq. \ref{eq:useful1}:
\begin{equation}
    \dot{\phi} \frac{1}{\sqrt{I}} \sum_i \boldsymbol{r}_{i}^{\perp} \cdot \boldsymbol{\dot{X}} +
    \dot{\phi} \frac{1}{\sqrt{I}} \sum_i \boldsymbol{r}_{i}^{\perp} \cdot \boldsymbol{r}_{i}^{\perp}
    = \frac{1}{\sqrt{I}} \sum_i \boldsymbol{r}_{i}^{\perp} \cdot \boldsymbol{\hat{n}}_i,
\end{equation}
then:
\begin{equation}
    \dot{\phi} = \frac{1}{I} \sum_i \boldsymbol{r}_{i}^{\perp} \cdot \boldsymbol{\hat{n}}_i,
\end{equation}
where we have used that $\sum_i \boldsymbol{r}_{i} = 0$.

As in the purely rotational case, we will utilize a rotating reference frame such that angles are defined with respect to $\phi$. The variables of interest evolve according to the following equations:
\begin{align}
    \dot{X} &= \frac{1}{N} \sum_i \boldsymbol{\hat{n}}_i \cdot \boldsymbol{\hat{x}}, \\
    \dot{Y} &= \frac{1}{N} \sum_i \boldsymbol{\hat{n}}_i \cdot \boldsymbol{\hat{y}}, \\
    \dot{\phi} &= \frac{1}{I} \sum_i \boldsymbol{r}_{i}^{\perp} \cdot \boldsymbol{\hat{n}}_i, \\
    \dot{\theta_i'} &= - \frac{1}{\pi_r} \frac{\partial V}{\partial \theta_i'} - \dot{\phi} + \sqrt{\frac{2 \pi_{\theta}}{\pi_r}} \xi_i,
\end{align}
with:
\begin{equation}
    V =- \frac{1}{2N} \sum_{ij} \left(\boldsymbol{\hat{n}}_i \cdot \boldsymbol{\hat{n}_j} \right) - \frac{1}{2I} \left(\sum_i \boldsymbol{r}^{\perp}_{i} \cdot \boldsymbol{\hat{n}}_i \right)^2.
\end{equation}

Then, $Q(X,Y,\phi,\theta_1',\dots,\theta_N' | t)$ evolves according to the following equation:
\begin{multline}
    \frac{\partial Q}{\partial t} = - \frac{\partial}{\partial X} \left[\frac{1}{N} \sum_i \boldsymbol{\hat{n}}_i \cdot \boldsymbol{\hat{x}} Q \right]
    - \frac{\partial}{\partial Y} \left[\frac{1}{N} \sum_i \boldsymbol{\hat{n}}_i \cdot \boldsymbol{\hat{y}} Q \right] \\
    - \frac{\partial}{\partial \phi} \left[\frac{1}{I} \sum_i \boldsymbol{r}_{i}^{\perp} \cdot \boldsymbol{\hat{n}}_i Q \right]
    + \sum_i \frac{\partial}{\partial \theta_i'} \left[\left(\frac{1}{\pi_r} \frac{\partial V}{\partial \theta_i'} + \dot{\phi} \right) Q \right]
    + \frac{\pi_{\theta}}{\pi_r} \sum_i \frac{\partial^2 Q}{\partial \theta_i'^2}.
\end{multline}

Let us define the reduced probability distribution $\mathcal{Q}(\theta_1', \dots, \theta_N' | t)$ such that:
\begin{equation}
    \mathcal{Q} = \int Q \, \mathrm{d} X \, \mathrm{d} Y \mathrm{d} \phi,
\end{equation}
and as neither $\frac{\partial{V}}{\partial \theta_i'}$ nor $\dot{\phi}$ depend on these variables, we are left with:
\begin{equation} \label{eq:fokker_planck_developed}
    \frac{\partial \mathcal{Q}}{\partial t} = \sum_i \frac{\partial}{\partial \theta_i'} 
    \left[ \left( \frac{1}{\pi_r} \frac{\partial V}{\partial \theta_i'} + \dot{\phi} \right) \mathcal{Q} \right]
    + \frac{\pi_{\theta}}{\pi_r} \sum_i \frac{\partial^2 \mathcal{Q}}{\partial \theta_i'^2}.
\end{equation}

We will now perform a scaling analysis to study the magnitudes of $\frac{\partial V}{\partial \theta_i'}$ and $\dot{\theta}$. Note the following:
\begin{align}
    \frac{\partial V}{\partial \theta_i'} &\sim -1 + R \dot{\phi}, \\
    \dot{\phi} &\sim \frac{1}{R},
\end{align}
where $R$ is the characteristic agent distance from the center of mass. If $\pi_r \ll R$, we may ignore the $\dot{\phi}$ term. More precisely, we can perform the same series expansion as in the purely rotational case, and arrive at an analogous result and higher-order corrections. Then, the Landau free energy in this case reads:
\begin{equation}
    f = \frac{1}{2} \left(\boldsymbol{\mu}_P^2 + \mu_{\phi}^2 \right) - \frac{1}{N \beta} \log (I_0(\beta D_i)),
\end{equation}
where:
\begin{align}
    \boldsymbol{\mu}_P &= \mu_x \boldsymbol{\hat{x}} + \mu_y \boldsymbol{\hat{y}} = \frac{1}{N} \left \langle \sum_i \boldsymbol{\hat{n}}_i \right \rangle, \\
    \mu_{\phi} &= \frac{I}{N} \langle \dot{\phi} \rangle,
\end{align}
and:
\begin{equation}
    D_i = \sqrt{N} \left \lVert \boldsymbol{\mu}_P + \sqrt{\frac{N}{I}} \boldsymbol{r}_{i}^{\perp} \mu_{\phi} \right \rVert.
\end{equation}

\subsection*{Zero-noise limit selection}

As the noise goes to zero, $\beta \rightarrow \infty$. We can then use the asymptotic expansion:
\begin{equation}
    I_0(x) \approx \exp(x),
\end{equation}
such that:
\begin{equation}
    f
    =
    \frac{1}{2} \sum_q \mu_q^2
    - \sum_i \left(\frac{1}{N} \sum_{q,l} \mu_q \mu_l \left(\boldsymbol{\varphi}^q_i \cdot \boldsymbol{\varphi}^l_i \right) \right)^{1/2}.
\end{equation}

Let us now minimize the functional $f$ in the presence of a single mode. We have the following:
\begin{equation}
    f_q = \frac{1}{2} \mu_q^2 - \frac{1}{\sqrt{N}} \sum_i \lVert \boldsymbol{\varphi}_i^q \rVert \mu_q,
\end{equation}
which has a minimum corresponding to:
\begin{equation}
    \widetilde{f}_q = - \frac{1}{N} \left(\sum_i \lVert \boldsymbol{\varphi}_i^q \rVert \right).
\end{equation}

We would like to maximize $\sum_i \lVert \boldsymbol{\varphi}_i^q \rVert$, subject to the normalization 
constraint $\sum_i \lVert \boldsymbol{\varphi}_i^q \rVert^2 = 1$. We can see that the maximum is achieved when
all the components are equal, i.e:
\begin{equation}
    \lVert \boldsymbol{\varphi}_i^q \rVert = \sqrt{\frac{1}{N}} \ \forall i.
\end{equation}

Then, in the $\beta \rightarrow \infty$ limit, collective motion will align with the translational mode, i.e. the
least localized mode. In the case of the ideal ring, because every particle is at the same distance, translations and rotations result in the exact same minimum and there is no preference of one over the other.

\section{Rotational-Auxetic system}

In the way that they have been previously defined, $\braket{\boldsymbol{\varphi}^{\phi} | \boldsymbol{\varphi}^{\gamma}} \neq 0$, which is not compatible with our formalism. We will use the
Gram-Schmidt formula to find an adequate orthonormal basis for the auxetic-rotational system. Starting with the rotational
mode, we can define an orthonormal mode to it as:
\begin{equation}
    \ket{\boldsymbol{\varphi}^B} = \frac{1}{\sqrt{1 - \braket{\boldsymbol{\varphi}^{\phi} | \boldsymbol{\varphi}^{\gamma}}^2}}
    \left(
    \ket{\boldsymbol{\varphi}^{\gamma}} - \braket{\boldsymbol{\varphi}^{\phi} | \boldsymbol{\varphi}^{\gamma}} \ket{\boldsymbol{\varphi}^{\phi}}
    \right).
\end{equation}
By defining:
\begin{equation}
    J = \sum_i \boldsymbol{r}_{i} \cdot \boldsymbol{\tilde{r}}_{i}.
\end{equation}

Then:
\begin{equation}
    \ket{\boldsymbol{\varphi}^B} = \frac{1}{\sqrt{1 - \frac{J^2}{I \widetilde{I}}}}
    \left(
    \ket{\boldsymbol{\varphi}^{\gamma}} - \frac{J}{\sqrt{I \widetilde{I}}} \ket{\boldsymbol{\varphi}^{\phi}}
    \right).
\end{equation}

For the sake of simplicity, we will rename $\ket{\boldsymbol{\varphi}^{\phi}} \rightarrow \ket{\boldsymbol{\varphi}^A}$.  Continuing with the dynamics of the system, we can decompose the motion of each particle as a rigid motion, rigid rotational motion, and auxetic deformation. Rigid motion is discarded as the system is pinned:
\begin{equation}
    \dot{\boldsymbol{x}}_{i} = \dot{\phi} \boldsymbol{r}_{c_i}^{\perp} 
    + \left(\dot{\phi}+\dot{\gamma}\right) \boldsymbol{\tilde{r}}_{i}^{\perp},
\end{equation}
where $\boldsymbol{r}_{c_i}$ is the position of the auxetic-circumference center of particle $i$. Note that $\gamma$ is defined with respect to the reference frame that rotates with $\phi$. With this, and using Eq. \ref{eq:useful1}:
\begin{align}
    \braket{\boldsymbol{\varphi}^A | \boldsymbol{\hat{n}}} &=  \braket{\boldsymbol{\varphi}^k|\dot{\boldsymbol{x}}}, \\
    &= \frac{1}{\sqrt{I}} \sum_i \left(\dot{\phi} \left(\boldsymbol{r}_{i}^{\perp} \cdot\boldsymbol{r}_{c_i}^{\perp} \right) 
    + \left( \dot{\phi} +\dot{\gamma} \right) \left(\boldsymbol{r}_{i}^{\perp} \cdot \boldsymbol{\tilde{r}}_{i}^{\perp} \right)\right),  \\
    &= \frac{1}{\sqrt{I}} \left(
     \dot{\phi} \sum_i \boldsymbol{r}_{i} \cdot \boldsymbol{r}_{c_i}
    + ( \dot{\phi}  + \dot{\gamma} ) \sum_i \boldsymbol{r}_{i} \cdot \widetilde{\boldsymbol{r}}_{i} 
    \right), \\
    &= \frac{1}{\sqrt{I}} \left( \dot{\phi} I + \dot{\gamma} J\right),
\end{align}
where we have used that $\boldsymbol{r}_{i} = \boldsymbol{r}_{c_i} + \widetilde{\boldsymbol{r}}_i$.

Similarly:
\begin{equation}
    \braket{\boldsymbol{\varphi}^B | \boldsymbol{\hat{n}}} = \frac{1}{\sqrt{1 - \frac{J^2}{I \widetilde{I}}}} 
    \left( \braket{\boldsymbol{\varphi}^{\gamma} | \dot{\boldsymbol{x}}}
    - \frac{J}{\sqrt{I \widetilde{I}}} \braket{\boldsymbol{\varphi}^{\phi} | \dot{\boldsymbol{x}}} \right).
\end{equation}

We will focus on the first term:
\begin{align}
    \braket{\boldsymbol{\varphi}^{\gamma} | \dot{\boldsymbol{x}}}
    &= \frac{1}{\sqrt{\widetilde{I}}} \sum_i \left(\dot{\phi} \left(\boldsymbol{\tilde{r}}_{i}^{\perp} \cdot\boldsymbol{r}_{c_i}^{\perp} \right) 
    + \left( \dot{\phi} +\dot{\gamma} \right) \left(\boldsymbol{\tilde{r}}_{i}^{\perp} \cdot \boldsymbol{\tilde{r}}_{i}^{\perp} \right)\right), \\
    &= \frac{1}{\sqrt{\widetilde{I}}} \sum_i \left(\dot{\phi} \left(\boldsymbol{\tilde{r}}_{i} \cdot\boldsymbol{r}_{c_i} \right) 
    + \left(\dot{\phi}+\dot{\gamma}\right) \left(\boldsymbol{\tilde{r}}_{i} \cdot \boldsymbol{\tilde{r}}_{i} \right)\right), \\
    &= \frac{1}{\sqrt{\widetilde{I}}} \left(\dot{\phi} J + \dot{\gamma} \widetilde{I} \right).
\end{align}

Then:
\begin{align}
    \braket{\boldsymbol{\varphi}^B | \boldsymbol{\hat{n}}} &= \frac{1}{\sqrt{1-\frac{J^2}{I \widetilde{I}}}}
    \left( \frac{1}{\sqrt{\widetilde{I}}} \left(\dot{\phi} J + \dot{\gamma} \widetilde{I} \right) - \frac{J}{\sqrt{I^2 \widetilde{I}}} \left( \dot{\phi}  I + \dot{\gamma} J
    \right)  \right), \\
    &=\frac{1}{\sqrt{1-\frac{J^2}{I \widetilde{I}}}} 
    \dot{\gamma} \left( \sqrt{\widetilde{I}} - \frac{J^2}{\sqrt{I^2 \widetilde{I}}} \right), \\
    &= \sqrt{1-\frac{J^2}{I \widetilde{I}}} \dot{\gamma}.
\end{align}

From this, it follows that:
\begin{align}
    \dot{\gamma} &= \frac{1}{\sqrt{\widetilde{I}}} \frac{1}{\sqrt{1 - \frac{J^2}{I \widetilde{I}}}} \braket{\boldsymbol{\varphi}^B | \boldsymbol{\hat{n}}}, \label{eq:gamma_evolution_auxrot} \\
    \dot{\phi} &= \frac{1}{\sqrt{I}} \braket{\boldsymbol{\varphi}^A | \boldsymbol{\hat{n}}} - \frac{J}{I} \frac{1}{\sqrt{\widetilde{I}}}
    \frac{1}{\sqrt{1 - \frac{J^2}{I \widetilde{I}}}} \braket{\boldsymbol{\varphi}^B | \boldsymbol{\hat{n}}}.
\end{align}

Now we can write the Fokker-Planck equation for $Q(\phi,\gamma,\theta_1',\dots,\theta_N')$, where $\theta_i' = \theta_i - \phi$ is the polarity angle of particle $i$ measured in the reference frame that rotates with $\phi$:
\begin{equation}
    \frac{\partial Q}{\partial t} = - \frac{\partial}{\partial \phi} \left[\dot{\phi} Q \right]
    - \frac{\partial}{\partial \gamma} \left[\dot{\gamma} Q \right]
    + \sum_i \frac{\partial}{\partial \theta_i'} \left[\left(\frac{1}{\pi_r} \frac{\partial V}{\partial \theta_i'} + \dot{\phi} \right) Q \right]
    + \frac{\pi_{\theta}}{\pi_r} \sum_i \frac{\partial^2 Q}{\partial \theta_i'^2},
\end{equation}
where:
\begin{equation}
    V = -\frac{1}{2} \left[\braket{\boldsymbol{\varphi}^A | \boldsymbol{\hat{n}}}^2 + \braket{\boldsymbol{\varphi}^B | \boldsymbol{\hat{n}}}^2  \right].
\end{equation}

In the rotating reference frame, there is no dependence on $\phi$, so we may integrate with respect to this variable for all orientations:
\begin{equation}
    \frac{\partial \mathcal{Q}}{\partial t} = 
    - \frac{\partial}{\partial \gamma} \left[\dot{\gamma} \mathcal{Q} \right]
    + \sum_i \frac{\partial}{\partial \theta_i'} \left[\left(\frac{1}{\pi_r} \frac{\partial V}{\partial \theta_i'} + \dot{\phi} \right) \mathcal{Q} \right]
    + \frac{\pi_{\theta}}{\pi_r} \sum_i \frac{\partial^2 \mathcal{Q}}{\partial \theta_i'^2}.
\end{equation}

We may recognize that as $\gamma$ evolves, the distances of the particles with respect to the pinning point change, and as such $I$ and $J$ are functions of this structure parameter. We will introduce the adiabatic approximation here
and assume that the system reaches a steady state configuration in a time such that $\gamma$ can be considered constant. Then, we are left with the following equation:
\begin{equation}
    \frac{\partial \mathcal{Q}}{\partial t} = 
    \sum_i \frac{\partial}{\partial \theta_i'} \left[\left(\frac{1}{\pi_r} \frac{\partial V}{\partial \theta_i'} + \dot{\phi} \right) \mathcal{Q} \right]
    + \frac{\pi_{\theta}}{\pi_r} \sum_i \frac{\partial^2 \mathcal{Q}}{\partial \theta_i'^2},
\end{equation}
which can be solved in the same way as before by invoking a series expansion, with the zeroth order solution corresponding to the Gibbs measure.

Then we can write the Landau free energy as:
\begin{equation}
    f = \frac{1}{2} \left(\mu_A^2 + \mu_B^2 \right) - \frac{1}{N\beta} \log \left(I_0 \left(\beta D_i \right) \right),
\end{equation}
where:
\begin{align}
    \mu_A &= \frac{1}{\sqrt{N}} \left \langle \braket{\boldsymbol{\varphi}^A | \boldsymbol{\hat{n}}} \right \rangle
    = \sqrt{\frac{I}{N}} \langle \dot{\phi} \rangle + \sqrt{\frac{J^2}{NI}} \langle \dot{\gamma} \rangle, \\
    \mu_B &= \frac{1}{\sqrt{N}} \left \langle \braket{\boldsymbol{\varphi}^B | \boldsymbol{\hat{n}}} \right \rangle
    = \sqrt{\frac{\widetilde{I}}{N}} \sqrt{1 - \frac{J^2}{I \widetilde{I}}} \langle \dot{\gamma} \rangle,
\end{align}
and:
\begin{equation}
    D_i = \sqrt{N}
    \left \lVert \mu_A \boldsymbol{\varphi}^A_i + \mu_{B} \boldsymbol{\varphi}^B_i \right \rVert.
\end{equation}

The only relevant structure parameter in our case is the auxetic angle $\gamma$, which will evolve adiabatically according to the ensemble average of Eq. \ref{eq:gamma_evolution_auxrot}:
\begin{align}
    \frac{\mathrm d \gamma}{\mathrm{d} t} &= \langle \dot{\gamma} \rangle, \\
    &= \sqrt{\frac{N}{\widetilde{I}}} \frac{1}{\sqrt{1 - \frac{J^2}{I \widetilde{I}}}} \mu_{B},
\end{align}
which defines the operator $L$ for the auxetic angle.

\section{Numerical scheme}

Given the stochastic nature of our problem, we simulated our equations using the Euler-Maruyama method:
\begin{align}
    \ket{\boldsymbol{x}}^{n+1} &= \ket{\boldsymbol{x}}^n + \ket{\boldsymbol{\hat{n}}}^{n} \Delta t + \ket{\boldsymbol{F}}^n \Delta t, \\
    \ket{\theta}^{n+1} &= \ket{\theta}^n + \frac{1}{\pi_r} \mathcal{K}^n \ket{\boldsymbol{F}}^n \Delta t +
    \sqrt{\frac{2 \pi_\theta}{\pi_r}} \ket{\xi}^n \sqrt{\Delta t},
\end{align}
where $\Delta t = 2 \times 10^{-4}$. Tensions were computed from Eq. \ref{eq:SSt_equation} using the biconjugated gradient stabilized routine in Eigen, a C++ library. Due to the presence of zero modes, these tensions may not be uniquely determined. But after multiplying by $\mathcal{S}^T$ to obtain the net force, the component in the kernel of $\mathcal{S}^T$ is eliminated and we get a net force that is unique.

The Euler-Maruyama scheme introduces errors in the position of the particles, which ultimately alters the structure of the system and the modes. This is more pronounced when dealing with the rotation of the rigid bars, which will lead to an increase in their length for any finite $\Delta t$. As such, a ``thermostat'' must be implemented. We put in place a simple routine that periodically calculates the structure parameters from the positions of the particles, which could be slightly different for different groups of particles due to the aforementioned infinitesimal deformations. After a regular timestep interval, the thermostat reconstructs the system using the space-averaged structure parameters. All these elements were implemented in a custom-made C++ code.

Free energies were minimized using the FindRoot routine in Mathematica. Depending on the modes present in the system, these energy landscapes can be complex and present various minima, sometimes very close to each other, so special care was taken to ensure that we found all the minima, and that we could follow each one individually in time if there is a structure parameter that is evolving. Such as in the rotational-auxetic system, where there are four minima a minimum is found using the routine. Then, the structure parameter is evolved using a forward Euler algorithm, which in the auxetic case reads:
\begin{equation}
    \gamma^{n+1} = \gamma^{n} + \langle \dot{\gamma} \rangle^n \Delta t.
\end{equation}
This procedure was repeated for various values of $\pi_{\theta}$ to obtain the phase diagram.